\documentclass[12pt]{article}
\usepackage{multicol}
\usepackage{color}
\usepackage{xspace}
\usepackage{enumerate}
\usepackage{dirtytalk}
\usepackage{textcomp}
\usepackage{natbib}
\usepackage{amssymb}
\usepackage{graphicx,epsf,amsmath,amsfonts,wrapfig}

\newcommand{\Msol}{\hbox{${\rm ~ M_{\odot}}$}}

\def\kms{kms$^{\rm -1}$}
\usepackage{authblk}
\renewenvironment{abstract}
 {\small
  \begin{center}
  \bfseries \abstractname\vspace{-.5em}\vspace{0pt}
  \end{center}
  \list{}{%
    \setlength{\leftmargin}{2mm}
    \setlength{\rightmargin}{\leftmargin}%
  }%
  \item\relax}
 {\endlist}
 
\begin{document}

\title{Jets from Young Stars\footnote{To appear in the New Astronomy Reviews special volume {\em 100 Years of Jets} (eds.\ Rob Fender and Ralph Wijers)}}

\author[1]{T.P.\ Ray}
\author[2]{J.\ Ferreira}
\affil[1]{ Dublin Institute for Advanced Studies, Ireland}
\affil[2]{ Universit{\'e} Grenoble Alpes, France}
\date{August, 2020}

\maketitle
\begin{abstract}
Jets are ubiquitous in the Universe and, as demonstrated by this volume, are seen
from a large number of astrophysical objects including active galactic nuclei, gamma
ray bursters, micro-quasars, proto-planetary nebulae, young stars and even brown
dwarfs. In every case they seem to be accompanied by an accretion disk and, while
the detailed physics may change, it has been suggested that the same basic
mechanism is responsible for generating the jet. Although we do not understand what
that mechanism is, or even if it is universal, it is thought to involve the centrifugal
ejection of matter from the disk along magnetic field lines.
For a number of reasons, in particular their proximity and the abundant range of
diagnostics to determine their characteristics, jets from young stars and their
associated outflows may offer us the best opportunity to discover how jets are
generated and the nature of the link between outflows and their accretion disks.
Recently it has become clear that jets may be fundamental to the star formation
process in removing angular momentum from the surrounding protoplanetary disk
thereby allowing accretion to proceed. Moreover, with the realisation that planetary
formation begins much earlier than previously thought, jets may also help forge planets
by determining initial environmental characteristics. This seems to be particularly true
within the so-called terrestrial planet forming zone.
Here we review observations of jets from young stars which have greatly benefitted
from new facilities such as ALMA, space observatories like Spitzer, Herschel and HST,
and radio facilities like LOFAR and the VLA. Interferometers such as CHARA and
GRAVITY are starting to make inroads into resolving how they are launched, and we
can look forward to a bright future in our understanding of this phenomenon when
JWST and the SKA come on stream. In addition, we examine the various
magnetohydrodynamic models for how jets from young stars are thought to be
generated and how observations may help us select between these various options.
\end{abstract}
\section{Historical Introduction}

In the 1950s, George Herbig and Guillermo Haro  \citep{1951ApJ...113..697H, 1952ApJ...115..572H} discovered the nebulous patches in Orion that now bear their name. At first these were thought to be actual \say{cradles} of stellar birth and it was not until the 1960s that Richard Schwartz \citep{1977ApJ...212L..25S} realised that Herbig-Haro (HH) objects may be part of an outflow from a young star. This he thought likely because their spectra bore certain similarities to the shock ejecta from supernova remnants. 

It was not however until the development of the Charge-Couple Device (CCD), with its vastly superior quantum efficiency in comparison to photographic plates, that the \say
{dots were joined}. The nebulous clouds seen by Herbig and Haro were found to be driven by, or parts of, a highly collimated line emitting jet.
At the same time the use of CCDs was coming to the fore, millimeter wave astronomy was undergoing a revolution. The first large millimeter grade dishes, in combination with  superheterodyne receivers, started to map star formation regions in single molecular transitions such as the fundamental 2.3 mm J = 1$\rightarrow$0 rotational line of CO. Hoping to see matter collapsing inwards to form a star, to their surprise the early pioneers discovered matter was actually flowing away: outflow rather than inflow was observed. Often diametrically opposite red-shifted and blue-shifted lobes were seen moving at velocities of tens of kms$^{\rm -1}$ and straddling the young star \citep{1980ApJ...239L..17S}. Since this gas had to be at very low temperatures, otherwise the 2.3\,mm CO transition would not be excited, such molecular outflows had to be highly supersonic. 

In a number of cases clear associations were seen between the optical outflows and their molecular counterparts. The optical line emission was observed to be strewn along the same axis as the molecular emission and the latter to encase the former \citep[e.g,][]{1985ApJ...290..587S}. As the area that could be surveyed by CCDs grew, with their increasingly larger format, it became clear that outflows from young stars could extend up to several parsecs, stretching, in some cases, beyond the boundaries of the young star's associated molecular cloud \citep{2012AJ....144..143B,2019ApJ...871..141Q}. 

As outflows were observed from a whole host of young stars of different masses and evolutionary phases, certain patterns were seen to emerge. First of all whether an outflow appeared or not depended on whether the young star was surrounded by an accretion disk. No accretion disk implied no jet, although the converse was not necessarily true. It was also found that the strength of an outflow depended on the evolutionary status of the parent star: less evolved stars have more powerful outflows. As we shall see, this is in line with the fact that such stars have stronger accretion rates, all else being equal, and the relationship between outflow and accretion rates (see \ref{Link-Accretion-Outflow}). 

At this point it may be useful to point out how we distinguish between young stars at different evolutionary phases. The standard picture, for solar-like stars, is that their formation begins through the radial collapse of a molecular core, perhaps a parsec across,    and that this gives rise to the birth of one or more stars \citep[e.g][]{2015arXiv151103457K}. In the earliest phase, most of the mass is still contained in the surrounding envelope, the energy output is entirely dominated by accretion onto the central object, and there are many tens of magnitude of visual extinction present. The source at this stage is referred to as a Class 0 \citep{1993ApJ...406..122A} young stellar object (YSO). Observations shows that such objects already possesses a disk as well as an outflow. The spectral energy distribution (SED) of a Class 0 YSO closely resembles a blackbody with temperatures of at most a few tens of K and a peak flux in the millimeter regime. The SED is dominated by the optically thick envelope. 

In the next phase, Class I \citep{1984ApJ...287..610L}, the protostar has either accreted or driven away enough of its envelope that the effects of extinction are considerably reduced. The YSO now emits in the near-infrared (NIR) and its SED is dominated by the contribution from the disk along with one from the envelope.  Once the YSO becomes optically visible, its photosphere can be observed, and it is now a T Tauri star (TTS) or, if a few times more massive than the Sun, a Herbig Ae/Be star. T Tauri stars are further sub-divided into those with accretion disks, so-called classical T Tauri stars (cTTS), or those without disks, referred to as weak-line T Tauri Stars (wTTS). These are also known as Class II and Class III YSOs respectively \citep{1984ApJ...287..610L}. Although estimates vary somewhat as to how long an individual young star takes to go through each phase, as a rule of thumb both cTTSs and wTTSs are at most a few million years old. At all stages of the process, the luminosity of the YSO is derived either from accretion or, in more evolved sources, from gravitational contraction. At no time has fusion been ignited in the core. As pointed out previously, jets are only present when a YSO is surrounded by an accretion disk, thus while a number of cTTS have been found to possess jets, jets are not found around {\em bona fide} wTTS.

\section{Physical Characteristics of YSO Jets}

Jets from young stars have been observed at a variety of wavelengths from the X-ray regime right down to the radio band. In this review we will concentrate on the optical and NIR bands as historically this is where much of the effort on understanding their nature has focused.   

\begin{figure}
\centering
\includegraphics[scale=0.18]{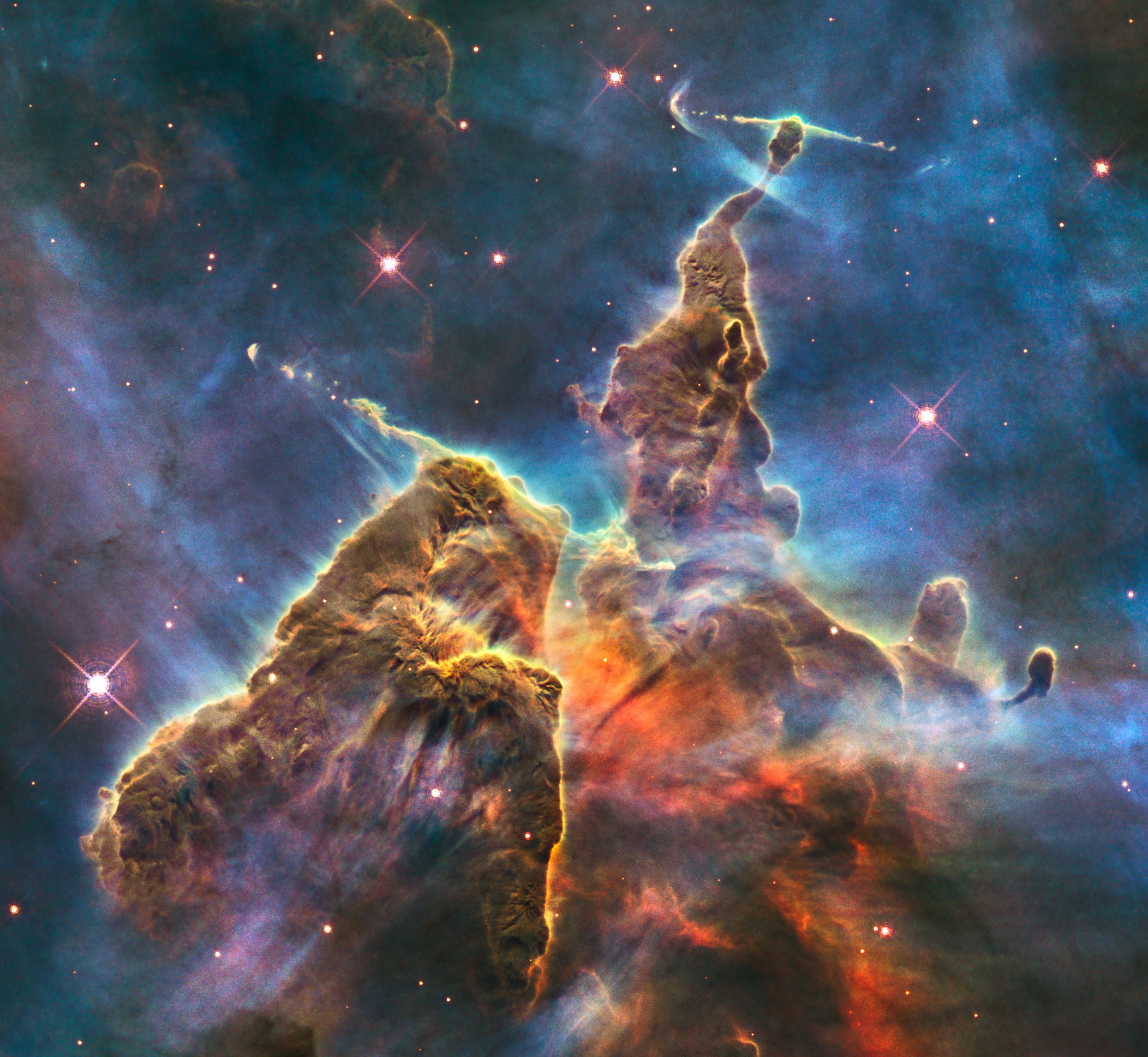}
\caption{The HH\,901 and HH\,902 outflows as imaged by HST's Wide Field Camera 3. The colours in this composite image correspond [OI] (blue), H$\alpha$ (green), and [SII] (red). Credit: NASA, ESA, M.\ Livio and the Hubble 20$^{\rm th}$ Anniversary Team (STScI).}
\label{Carina_Jets}
\end{figure}

The radiation from YSO jets is for the most part characterised by line emission, at least from the UV to the sub-millimeter bands. At optical and NIR wavelengths, this is a combination of both permitted, e.g.\ H$\alpha$, and forbidden lines, e.g\ [OI]$\lambda$630\,nm, 
[SII]$\lambda\lambda$671.6\,nm, 673.1\,nm, [FeII]$\lambda$1.64\,$\mu$m, etc.\ (see Fig. \ref{Carina_Jets}). The presence of such forbidden lines immediately implies that electron densities must be low (at most 10$^{\rm 4\mbox{--}5}$cm$^{\rm -3}$) in some parts of the jet \citep{1995ApJ...452..736H}. Moreover, as also recognised by \cite{1977ApJ...212L..25S} many years ago, the line emission in Herbig-Haro jets comes from the cooling zone of shocks, rather than from photo-ionization.

This presence of line emission can thus be used to infer fundamental parameters such as a jet's radial velocity, and typical values of around a few hundred \kms are found. We expect of course tangential velocities to be similar and this is consistent with observed proper motions \citep[e.g.,][]{1992A&A...263..292E}. In fact, the combination of radial velocities, proper motions and known distance to the source, gives us a full 3-D kinematic picture of a jet. Moreover the inclination angle of the outflow to the plane of the sky can be determined in this way. 

Now line spectra can also be used to derive a jet's temperature and therefore its local sound speed. As the temperatures found are around 10$^{4}$\,K, i.e.\ the sound speed is approximately 10\,\kms, YSO jets are thus highly supersonic with Mach numbers, M$_{\rm Jet}$, of around 20-30. Since the opening angle (in radians) of an unconfined supersonic flow $\approx \frac{\rm 1}{\rm M_{Jet}}$, i.e. the sound speed divided by the jet's velocity, we expect values of a few degrees at most. Since these are the values typically found not only in atomic but even in some molecular species \citep{2007A&A...462L..53C}, we can assume jets for the most part are advancing ballistically or at least they are not confined laterally. Note that this only refers to distances of several hundreds of au from the source where the magnetic field is expected to be too weak to play a role in the jet's dynamics \citep{2007ApJ...661..910H}. Closer to the source, magnetic fields are expected to play a major role in launching and focusing the jet. 

\cite{1992ApJ...386..222R} noted the morphological similarity of YSO jets, such as the HH\,34 system \citep{1988A&A...200...99B}, to those from Active Galacitc Nuclei (AGNs) and in particular they proposed a similar model to explain their knots, as suggested by \cite{1978MNRAS.184P..61R} for the M87 jet, i.e.\ the observed knots are so-called \say{working surfaces} i.e.\ internal shocks.  Such shocks occur as a result of velocity variations in a highly supersonic flow as more rapidly advancing jet material catches up with slower material ahead of it. The effective shock velocity is then determined by the {\em difference} in velocity of material rather than by its absolute value. The latter would only be relevant for a light jet flowing into a relatively dense stationary medium, e.g.\ molecular cloud material or the interstellar medium (ISM) at the terminus of a jet. The existence of working surfaces explains a conundrum: why the indicated shock velocities at knots are often so low (typically a few tens of \kms in comparison to their bulk speeds \citep{1994ApJ...436..125H}.

That is not to say jet shock speeds are always low: in fact, the largest scale structures which resemble giant bows, such as those in HH\,34  and HH\,212  are associated with very high shock velocities. This is implied, not only through their spectroscopic line profiles but also by the high degree of ionisation at their apex, giving rise to, for example, [OIII] emission \citep{1988A&A...200...99B}. 

Of course shocks in YSO jets and those in AGN jets differ in a fundamental way apart from the obvious dissimilarities in size and velocity. Shocks in AGN jets are essentially adiabatic, in relative terms the energy losses at their working surfaces are small. In contrast, shocks in YSO jets are radiative, and sometimes a significant fraction of the shock energy can be converted into line emission. 

From a diagnostic perspective, and as already mentioned, the presence of line emission offers a distinct advantage over the broad, often bland, continuum seen in AGN jets: using line diagnostics one can readily determine a jet's fundamental parameters such as temperature, velocity, electron density, etc. Moreover, indirectly we can obtain others. For example, using the assumption of charge exchange, i.e.\ species such as sulphur or oxygen become ionized, or end up in an excited neutral state, through charge exchange with hydrogen, the total density of neutral plus ionized species can be obtained. Modeling then shows \citep[e.g.,][]{1999A&A...342..717B} that YSO jets are mostly neutral with ionisation levels typically around 1--10\% \citep{2009A&A...506..779P}. 

If we have the total density then clearly using the jet radius, in combination with its velocity, we can determine the mass flux through the jet. Such fluxes are found to vary widely (10$^{\rm -5}$--10$^{-8}$\,\Msol/yr) but in general a number of expected trends are seen.  For example, mass loss rates are observed to decrease with advancing evolutionary phase for a young star of fixed mass but also increase with the mass of the protostar \citep{2016MNRAS.460.1039P}. Since outflow rates roughly correlate with accretion rates, this reflects what is seen for mass accretion rates onto young stars \citep[e.g.,][]{2014A&A...572A..62A}. 

\section{What an Outflow from a Young Star Looks Like: The Case of HH~211}\label{Case of HH211}
\begin{figure}
\centering
\includegraphics[scale=0.27]{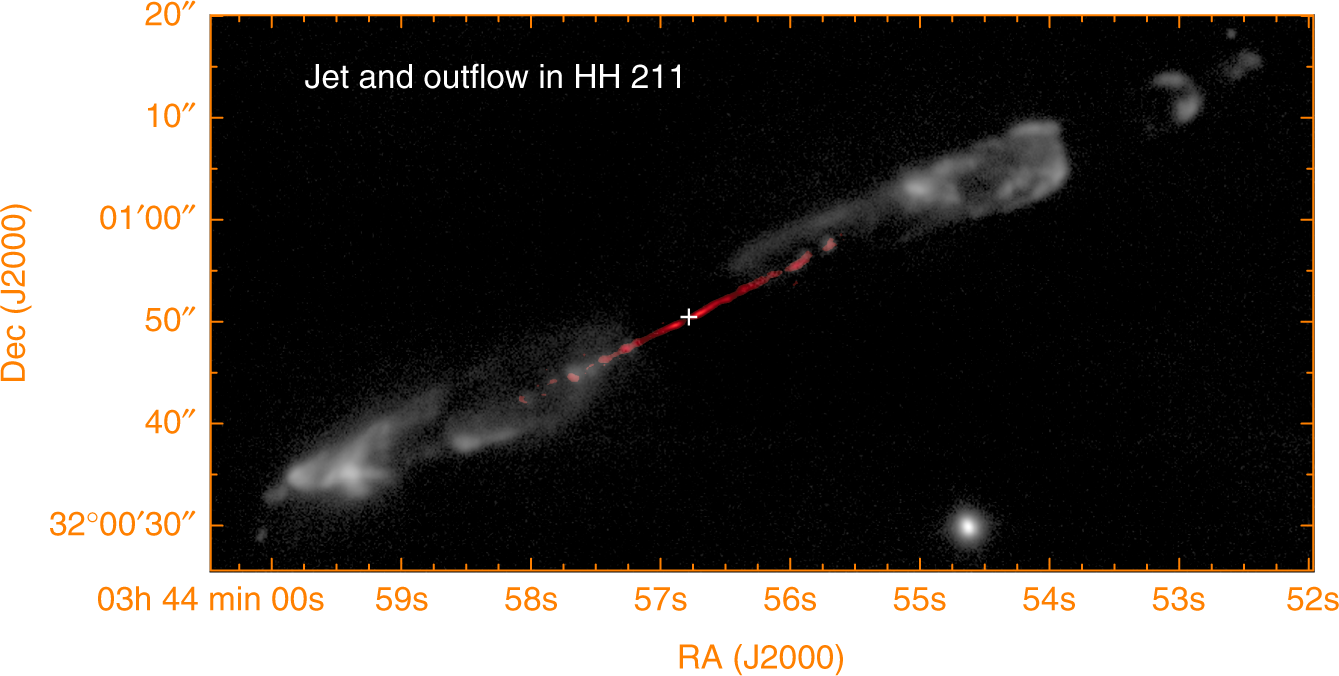}
\caption{The HH~211 Outflow near IC~348 in shocked molecular hydrogen emission at 2.12$\mu$m. The central source is a highly embedded Class 0 YSO the position of which is marked by a white cross. Red shows shocked SiO emission. This is in the form of a highly collimated jet that appears to be driving the molecular outflow. Credit: \citet{2018NatCo...9.4636L} and reproduced with permission of the lead author.}
\label{HH211}
\end{figure}

HH\,211 was the first outflow from a young star that was found, not in the optical but, in the NIR using a narrow band filter centered on the 2.12$\mu$m line of shocked molecular hydrogen \citep{1994ApJ...436L.189M}. Two clear large bow shocked-shaped features are seen pointing to the northwest and southeast of its source (marked with a white cross in Figure \ref{HH211}). It is close to the young stellar cluster IC\,348 and is located approximately 300\,pc away. The source itself is a highly embedded Class~0 YSO and the outflow appears to be very young: a dynamical timescale of only 1,000 years is inferred from dividing its projected length by its tangential velocity of approximately 100 \kms \citep{2016ApJ...816...32J}.  Following on from its discovery in the NIR, molecular observations at longer wavelengths, using the CO J=1-0 and J=2-1 lines as tracers, also revealed a very collimated bipolar structure \citet{1999A&A...343..571G}. In fact if we assume the bow features observed in molecular hydrogen emission are where the bipolar jet rams into and shocks the surrounding interstellar medium, then the morphology of the CO emission clearly suggests 
cavities in the wake of these terminal shocks. This is in line with shock-entrainment as the formation mechanisms for young, embedded molecular outflows \citep{1999A&A...343..571G}. 

In other molecules, such as SiO, the outflow appears much more highly collimated than in low rotational transitions of CO (compare Figs.\ \ref{HH211} and \ref{Case of HH211}). Species such as SiO are enhanced, by several orders of magnitude, through grain-sputtering in shocks \citep{2007A&A...462..163N}.  Thus in Fig.\ \ref{HH211}, where we see SiO emission, we are really observing where the shocks are strong enough to significantly impact on dust grains. Although the HH\,211 region is clearly highly embedded, and thus difficult to observe optically, nevertheless the presence of an underlying atomic jet can be inferred from far-infrared observations, e.g.\ [OI]\,63$\mu$ and 145.5$\mu$ data from Herschel/PACS \citep{2018A&A...616A..84D}.

\begin{figure}
\centering
\includegraphics[scale=0.2]{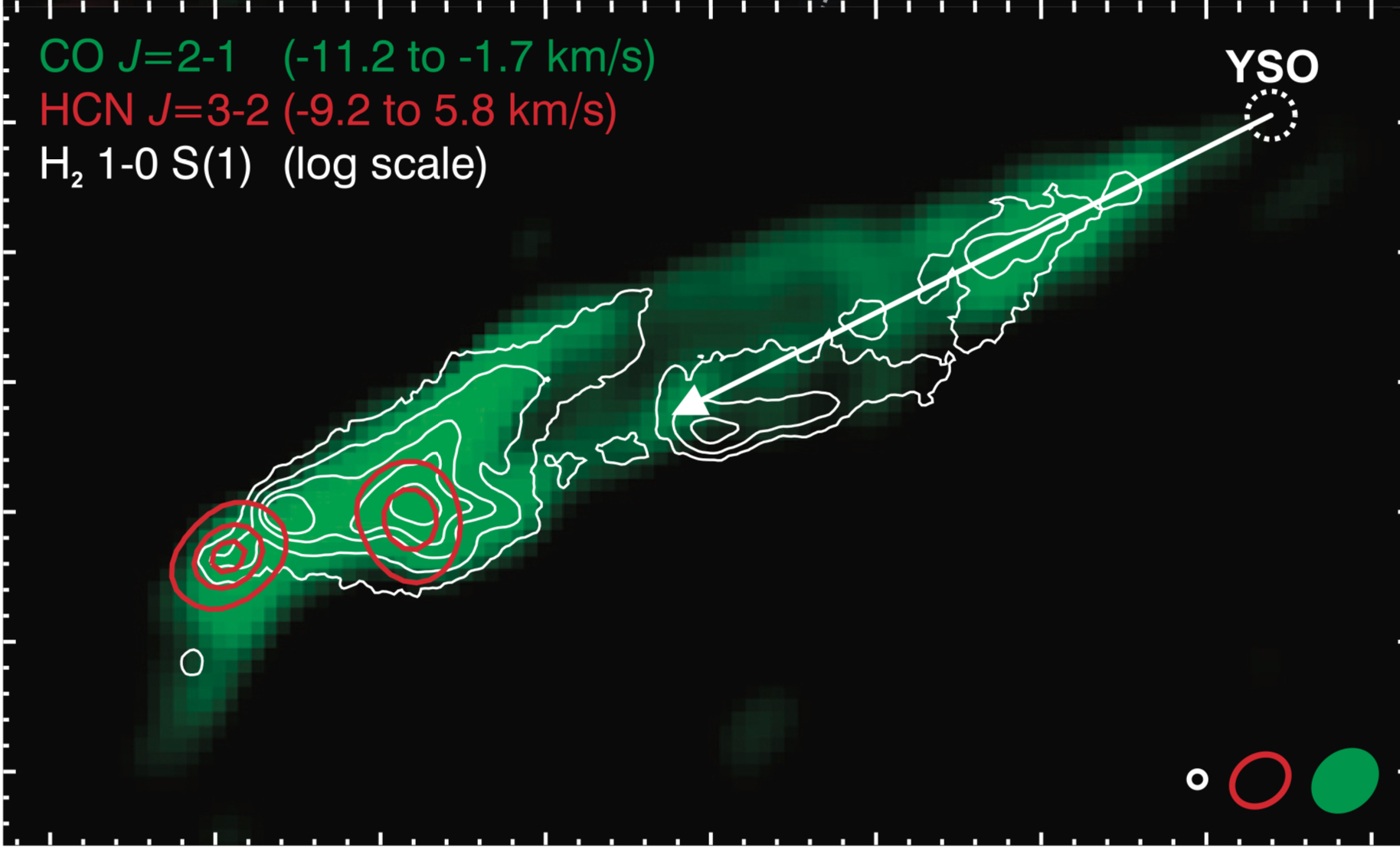}
\caption{A comparison of NIR H$_{\rm 2}$ 2.12$\mu$m (white contours) emission and CO J=2$\rightarrow$0 emission (in green) for the HH\,211 blueshifted SE lobe. The CO emission, which covers the velocity interval of -11.2 to =1.7 \kms\, here, was mapped using the SMA. Some HCN J=3$\rightarrow$2 knots near the apex are also seen. At more blueshifted velocities the CO emission is closer to the axis of the SiO jet observed in Fig.\ \ref{HH211}.  Effective beam sizes are shown in the bottom right corner. Credit: \citet{2012ApJ...751....9T}. Reproduced with permission of the authors and \textcopyright AAS.}
\label{Tappe2012}
\end{figure}

\section{What Radio Emission from Outflows Has to Tell Us}

Early observations of jets from young stars with, for example, the Very Large Array (VLA) showed them to be faint radio emitters with the flux at frequency $\nu$, S$_\nu \lesssim$ 1 mJy at around 5\,GHz). Their spectra were found to be thermal with S$_\nu \propto \nu^{\alpha}$ and $\alpha$ in the range 0.1 to 1.0 \citep{2018A&ARv..26....3A}.  Models approximating the radio emission as a wind of fixed opening angle (typically a few tens of degree) appeared to fit the data well \citep{1986ApJ...304..713R}. Maps of the source often showed the emission to be elongated in the direction of the known optical/near-infrared jet and occasionally more distant radio emission was seen, again along the jet axis. In a number of cases, proper motion measurements could be made and these were found to be close to those derived from line emission at optical/near-infrared wavelengths, i.e.\  a few hundred \kms\ \citep[e.g.,][]{1989ApJ...346L..85R}. Assuming total (radial plus tangential) jet velocities to be similar to those of their optical counterparts, observed radio fluxes can be used in combination with known distances to derive mass loss rates. While these were typically found to be considerably lower than those inferred from optical line emission, the different values can be reconciled by noting that the latter represents both the atomic (neutral) and ionized component whereas the radio emission only comes from the ionized component. Thus mass loss rates derived from radio emission are effectively lower bounds and typically suggest 1-10\% of a jet is ionized. 

Since however the early observations of YSO jets described above, the sensitivity of  interferometers such as the VLA in the US or MERLIN in the UK have increased enormously. This is as a result of the broader bands now observable using fibre optic links\footnote{In the past microwave links or coaxial cable were used with very limited bandwidth.}. For example observations carried out in so-called C-band (from 4-8\,GHz) routinely encompass 1 GHz either side of the chosen central frequency. Such an increase in sensitivity is leading to a resurgence of interest in mapping YSO outflows and even carrying out large surveys of star formation regions \citep{2018ApJS..238...19T}. Moreover the availability of very low frequency arrays such as LOFAR are beginning to make an impact in this field. For example in the case of thermal emitting jets, the expected turnover at low radio frequencies has been found thus allowing us an independent estimate of the electron density and the total mass of ionized gas \citep{2017ApJ...834..206C}. 

Perhaps however the most intriguing finding in the past few years is that a growing number of outflows have non-thermal emission or, at least, a non-thermal component mixed in with thermal one. The classic example is the HH80/81 outflow associated with the massive young star IRAS 18162-2048. This jet stretches over 5\,pc and while the radio spectral index towards the centre of the outflow suggests thermal emission, a non-thermal index is derived further out. Moreover, the jet has been observed to be linearly polarized, a clear signature of synchrotron emission \citep{2010Sci...330.1209C}. Soft X-ray thermal emission has been observed ahead of the non-thermal radio peak \citep{2019MNRAS.482.4687R} suggesting a forward shock that heats gas up to 10$^{\rm 6}$\,K and a reverse shock (Mach disk) where the relativistic particles are likely to be accelerated. Other cases of non-thermal emission have been found \citep[e.g.,][]{2014ApJ...792L..18A} in outflows from solar-like young stars where the velocity of the outflow is considerably lower than the 1,000 \kms reached in HH80/81. Clearly, and in conditions very unlike those seen in supernova remnants for example, it is possible to accelerate particles to high energies. How this is done is still under debate \citep[e.g.,][]{2010A&A...511A...8B} but an interesting prediction of such models is that YSO jets should produce $\gamma$-ray emission through the Inverse Compton process. Recent observations, based on a time base of 10 years with the Fermi-LAT satellite, show the HH80/81 jet may well be a source of $\gamma$-ray photons up to 1 GeV \citep{2019arXiv190810994Y}. 

Finally it should be remarked that low frequency observations of non-thermal emission can offer us another method of measuring the magnetic field in an outflow. This technique relies on the fact that an outflow lobe can consist of a mixture of relativistic particles in combination with thermal electrons. The latter, depending on the electron density, de-collimates the forward-beamed radiation from the non-thermal electrons, giving rise to a decrease in flux or effectively a turnover in the synchrotron spectrum. If the electron density is known from line diagnostics (e.g.\ using the ratio of the red [SII] doublet) then the magnetic field can be derived. This, so-called Razin Effect, may have been seen in a YSO outflow for the first time \citep{2019ApJ...885L...7F}. The derived magnetic fields are in line with what is expected, and as found, by totally independent line diagnostic techniques \citep{2015ApJ...811...12H}. 

\section{Linking Accretion with Ejection}\label{Link-Accretion-Outflow}
Accretion onto a young star can be measured in a variety of ways \citep{2014A&A...572A..62A} and an obvious link to look for is between the level of accretion and the strength of its outflow. This is perhaps an appropriate juncture to explain in more precise terms what is meant by the term \say{outflow}. The typical velocities, i.e.\ the sum of the radial and tangential velocities, of material observed in optical/NIR lines for example, is a few hundred kms$^{-1}$, i.e.\ comparable to the escape velocity from the young star. In other words this component of the outflow clearly originated from deep in the gravitational well. In contrast molecular outflows, at least in low excitation lines such as the fundamental rotational transition of the CO molecule at 2.3\,mm, are seen to be slowly moving, i.e.\, roughly at a few tens of \kms. Despite its slow motion, the total mass of molecular gas set in motion is considerable. This was recognised very early on \citep[e.g.,][]{2006ApJ...649..280S} with the discovery that while the molecular outflow might consist of several solar masses of material, the star at the centre of the outflow could be less than a solar mass. The inference from this is either star formation is very inefficient in the sense that many solar masses of material must be funneled in and away from a protostar for the latter to acquire even a mass comparable to the sun or alternatively that the high velocity outflow from the young star is very efficient in driving molecular material around it. Modeling shows the latter statement to be more likely to be correct \citep{2000prpl.conf..867R}. In other words, to borrow a phrase from Theology used first by Saint Thomas Aquinas, the high velocity jet is, at least to a first approximation, the\say{prime mover}. 

\begin{figure}
\centering
\includegraphics[scale=0.12]{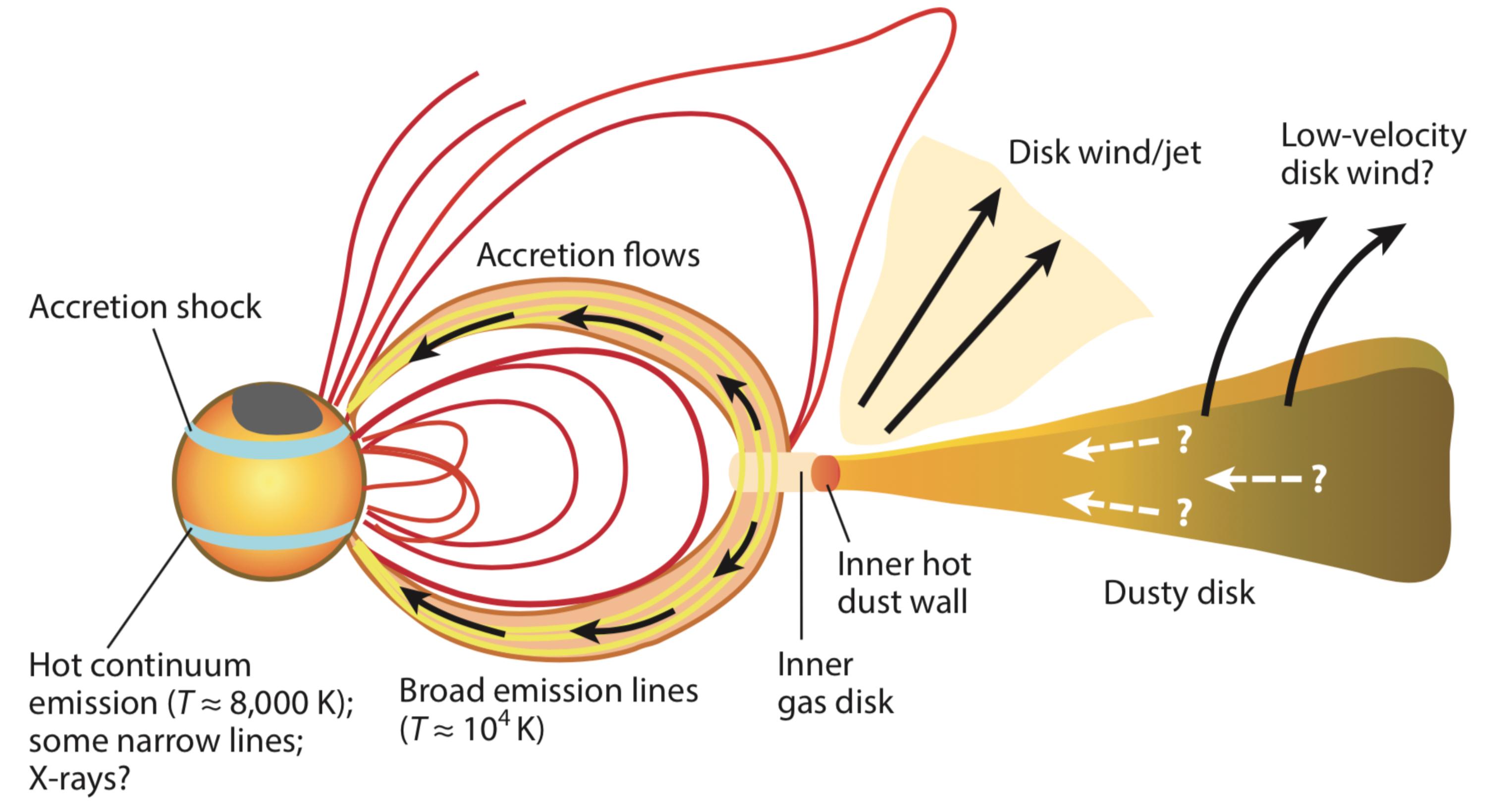}
\caption{Accretion and winds around YSOs. The standard paradigm is that accretion, and hence radially inwards motion, occurs in the first instance through the disk at least up until the so-called co-rotation radius. To do so material must lose angular momentum perhaps as a result of the wind possessing spin. From the vicinity of the co-rotation radius accretion then proceeds along magnetically guided columns onto the star. Before impacting on the stellar surface, the accreted material has to be slowed down through a shock. Emission from the columns is responsible for the time-varying, complex emission line profiles observed at the source.  Credit: \citet{2016ARA&A..54..135H}and reproduced with the permission of the lead author and \textcopyright ARA\&A.}
\label{Hartmann2016}
\end{figure}

Returning then to the question of the link between accretion and outflow: we can now define the outflow in a consistent way by looking at the atomic line emission. A number of authors have investigated this problem using both values for the outflow determined from extended, i.e.\ distant, spatially resolved sections of the jet \citep{1994ApJ...436..125H} as well as using line emission close to the source that is identified, through its kinematic and other physical properties, as the spatially unresolved jet component \citep{1995ApJ...452..736H}. Moreover, aside from optical and NIR lines,  mid- and far-infrared atomic lines can also be used as a diagnostic for more embedded sources and outflows \citep{2016ApJ...828...52W}. Estimating accretion onto a star can be done in a number of ways. All pre-main sequence stars derive their luminosity from either the release of potential energy through gravitational contraction and/or the depositing of matter onto the surface of the star (see Fig.\ \ref{Hartmann2016}),  along magnetic field lines that acts as funnels \citep{1997IAUS..182..433E}. The gravitational contraction timescale, also known as the Kelvin-Helmholtz timescale is relatively long, i.e.\ tens of millions of years for a solar mass star, whereas the accretion timescale is much shorter. In other words the accretion luminosity dominates over any contribution from contraction if the protostar is young enough. Thus the stellar luminosity: 
\begin{equation}
    L_{*} \approx GM_{*}\dot{M}_{acc}/R_{*}
    \end{equation}
is a good measure of accretion, $\dot{M}_{acc}$ asssuming the stellar mass, $M_{*}$, and the radius, $R_{*}$, is known. 
In the case of the optically visible classical T~Tauri stars, or their more massive counterparts the Herbig~Ae/Be stars, the line profiles from their photospheres in combination with a measure of their photospheric luminosities, allowing for extinction, can also be used as proxies to deduce an accurate accretion luminosity \citep{2014A&A...572A..62A}. An alternative approach, which produces similar results in the case of classical T~Tauri stars, is to measure their UV excess \citep{2015A&A...581A..66V}. This excess, not seen in the case of main sequence stars of similar spectral types, comes from radiation produced by material accreted onto the star as it goes through a strong shock close to the stellar surface \citep{1991ApJ...370L..39K} and see Fig.\ \ref{Hartmann2016}. In addition to the UV excess, there is a continuum from the accretion shock that extends into the optical. While the UV emission appears as a \say{bump} on the Wien part of the spectral energy distribution, the continuum extends into the photospheric spectrum of the star itself. This means that normal photospheric absorption lines are \say{filled-in}, a phenomenon known as line veiling \citep{1998ApJ...509..802C}. Veiling then can be used as an accretion measure. Finally it should be mentioned that a number of emission lines can be used effectively as proxies for accretion once the correlation between line luminosity, allowing of course for any extinction effects, and accretion luminosity is determined \citep[e.g.][]{2017A&A...600A..20A}.

Ultimately of course accretion (Fig.\ \ref{Hartmann2016}) is thought to occur through the disk surrounding the star. In order for disk material to move radially inwards, angular momentum must be removed from it first. Disk winds, which can subsequently be collimated into jets, can do this if they are ejected with spin. Alternatively viscosity in a disk can redistribute angular momentum by moving some material inwards, and allowing it to accrete, while at the same time transporting the excess angular momentum outwards. In this way disks actually grow with time, i.e.\ viscous spreading occurs. There is some evidence that this might occur, e.g.\ \cite{2018ApJ...864..168N} find that most Class II (T Tauri) disks are typically larger than those from an earlier evolutionary phase. There is, however, a large spread in sizes and  many Class II disks are relatively small. Ultimately disk sizes may be controlled through a combination of disks winds, photo-evaporation and the effects of newly formed planets. 
Models of disk accretion, i.e.\ the possible means to transport the disk angular momentum away, will be reviewed below, in the section on disk winds. 

\section{The different scenarios for YSO jets}

Discussions about the physical origin of YSO jets followed the same pattern as in AGN jets. Indeed, jets could either be powered by the central object (stellar winds) or linked to the circumstellar accretion disk (disk winds). Moreover, besides finding a model for jet production, one should also seek a mechanism regulating the spin evolution of a protostar. Indeed, it was soon realized that Class II T Tauri stars were slow rotators with typical periods of 8 to 10 days, i.e.\ only about 10\% of their breakup speed \citep{bouv86,bert89}. This is one of the yet unsolved mysteries of star formation: where does the stellar angular momentum go during the embedded phase (Classes 0 and I)?  
In addition, it appears that as long as the circumstellar disk is present (Class II), the protostar maintains a constant period, a situation referred to as $"$disk locking$"$ (see \citealt{gall13} and references therein). This is another mystery since the protostar is actually contracting and accreting material with large angular momentum so that it should actually spin up. Somehow, jets might also be an attractive way to get rid off the stellar angular momentum. Below, we review separately each class of models, since they co-existed in the literature until recently.    

\subsection{Stellar winds}

The first scenario invoked to explain the collimation of molecular outflows is based on a de Laval nozzle effect, due to the density gradient of the interstellar medium, acting on a stellar wind \citep{cant80}. This is the YSO equivalent of the \citet{blan74} twin exhaust model for AGN radio jets. However, it was also soon realized that YSO jets carry away far too much thrust and that they could not be radiatively driven \citep{deca81}. The other way to produce a stellar wind is to rely on magnetic acceleration. It was also known, since \citet{chan80} and \citet{blan82}, that large scale magnetic fields would not only accelerate loaded material, but also provide a hoop stress leading to self-confinement, conveniently taking care of the jet collimation issue. 

The idea of magnetic braking by a stellar wind is now very familiar \citep{mest87}. \citet{scha62} already pointed out that {\em $``$if gas emitted from a star is kept corotating with the star by the magnetic torques out to large distances, it will carry off far more angular momentum per unit mass than gas retaining the angular momentum of the stellar surface$"$}. This is the reason why using a magnetized stellar wind to spin down a protostar was one of the first ideas to come to mind. Unfortunately, the low rotation rate of T Tauri stars forbids any efficient centrifugal acceleration and some ad-hoc Alfv\'enic turbulent pressure must be invoked to lift material away from the protostar \citep{hart82a}. However, such a model had severe difficulties in explaining YSO jet mass losses and this idea has been discarded for quite some time.

\begin{figure}
\centering
\includegraphics[width=0.5\textwidth]{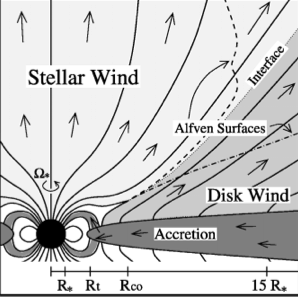}
\caption{Sketch of the magnetized star-disk interaction within the Accretion Powered Stellar Wind model. The stellar magnetic field truncates the disk at $R_t$, assumed to be located inside the corotation radius, $R_{co}$. The disk may have its own magnetic field, leading to the launching of an outer disk wind. Inside $R_t$, disk material is lifted towards the star but a fraction of it is ejected away.  Credit: \citet{matt05b}.}
\label{Matt+Pudritz05}
\end{figure}

The revival of stellar winds occurred with the Accretion Powered Stellar Wind (APSW) model of \citet{matt05b}. In this scenario, the stellar magnetic field (assumed dipolar) threads the accretion disk and leads to the formation of accretion funnel flows (see Fig.\ref{Matt+Pudritz05}). Because of magnetospheric stresses, the disk material is deflected away from the equatorial plane and falls towards the protostar along the dipolar field, leading to an accretion shock. The main assumption of the model is that a fraction of the mass and its associated accretion energy is actually deflected away along open polar stellar field lines, giving rise to an enhanced stellar wind. In principle, such a model is consistent with the observed accretion-ejection correlation and provides efficient magnetic braking. Allowing for protostellar contraction, \citet{matt12a} computed the evolution of stellar spin subject to an exponentially decreasing rate of accretion with time and showed that, for some parameters, it is  possible to achieve low rotation rates over typical TTS lifetimes (1 to 3 Myrs). 

This would require however mass loss rates in APSWs comparable to those inferred in YSO jets, implying that YSO jets are mostly made of material ejected from the protostar, in contradiction with jet kinematics \citep{ferr06b,cabr07}. Moreover, the model has a theoretical drawback summarized in the expression $"$you can't have your cake and eat it$"$. APSWs tap accretion energy, so the more massive they are the more energy they need. Each watt an APSW consumes, is a watt that is not being radiated away in the UV in the accretion shock (assuming a 100\% conversion efficiency). Since UV luminosity provides a proxy for the disk accretion rate, tapping this energy while maintaining the same UV emission implies that even more mass is actually being accreted onto the protostar. This enhances the spin up torque which, in turn, would require a larger spin down from the APSW. It can be proven that such a situation does not converge for some objects \citep{zann11}. 

APSWs are thus unlikely to be the main mechanism providing the required spin down torque for TTS. Although they cannot represent the main component of YSO jets, they probably fill in the jet axis, leading to some interaction with the surrounding jet medium (see eg \citealt{meli06b}).

\subsection{Outflows from the star-disk interface}

\begin{figure}
\centering
\includegraphics[width=0.85\textwidth]{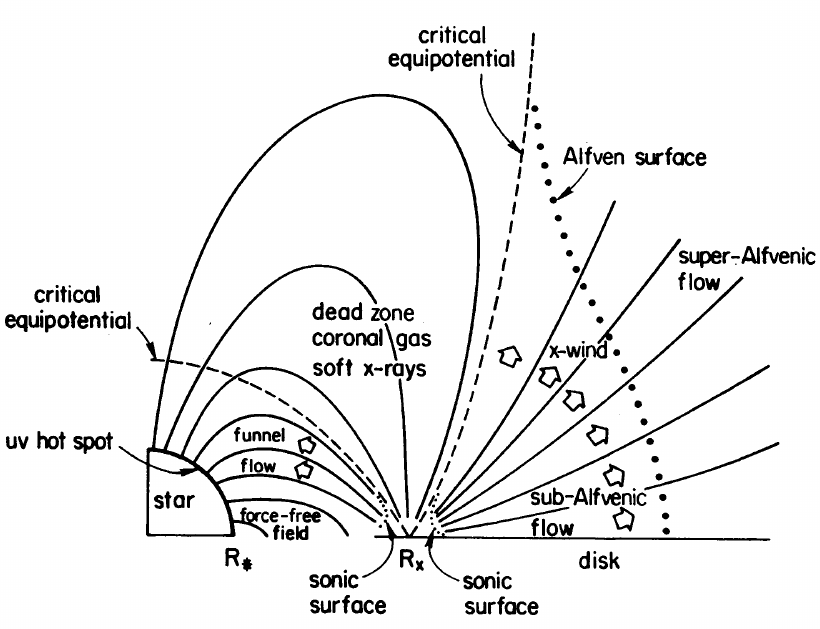}
\caption{Sketch of the X Wind in a typical Class II system. In this star-disk interaction model, the stellar magnetic field is the only relevant one, truncating the disk at a radius $R_X$ assumed to be the corotation radius. Stellar field lines below $R_X$ enforce corotation with some disk material, leading to infalling funnel flows. Field lines that would have thread the disk beyond $R_X$ are assumed to have either open up and disconnect from the star, leading to some highly localized wind from the disk (X-wind), or have inflated and remain still connected to the star (massless magnetized corona). Credit: \citet{shu94a}.}
\label{Shu94b}
\end{figure}

In 1988, F. Shu and collaborators proposed the X-celerator mechanism for YSO jets, where the protostar is assumed to rotate at break up speeds \citep{shu88}. The double goal sought by this scenario is to explain YSO jets while simultaneously braking the spin of the protostar. The disk material would reach the star at the equatorial plane and enforce the dipolar stellar magnetic field to open up, leading to centrifugal ejection from the disk mid-plane. But this early model was put aside as it became clear that TTS only rotate at typically 10\% of their break up speed. 

The X-wind scenario, as put forward by \citet{shu94a} with its various extensions (eg \citealt{cai08} and references therein), is an enhanced version of the same idea (see Fig.\ref{Shu94b}). More precisely, X-winds should be the dominant component in YSO jets (in terms of appearance, mass flux and power) while carrying away enough angular momentum so that the star is not being spun up by the accreting material (zero net torque condition, protostellar contraction neglected). In the X-wind paradigm, the stellar magnetic field is the only relevant large scale magnetic field present in the system. It truncates the accretion disk at the corotation radius\footnote{The corotation radius is the cylindrical distance where the stellar rotation rate is equal to the local Keplerian rotation rate.} leading to the formation of accretion funnel flows. However, a fraction of the stellar magnetic field is assumed to open-up and diffuse beyond the co-rotation radius. Then, from what is assumed to remain a point-like source around the co-rotation radius, a magnetized wind is expected to form and carry away all the angular momentum of the accreting material and a significant fraction of the incoming mass. 

Because of its elegance, this steady-state jet model had a huge impact on the YSO community. However, it turned out to suffer several severe drawbacks. First, YSO jet kinematics were shown to be inconsistent with published X-wind calculations. For instance, the observed range and steep radial decline of poloidal speeds could not be explained with the chosen Alfv\'en surface shape \citep{ferr06b, cabr07}. Second, the conditions required on the underlying disk (used as a boundary condition) turned to be quite extreme. According to this scenario, material accreted onto the star via funnel flows leaves its angular momentum at the base of the funnels, i.e.\ inside the co-rotation radius. In order for the X-wind to carry away all the excess angular momentum, some unknown outward (viscous) transport process must be invoked. This is rather unrealistic and probably explains why X-winds have never been obtained so far in MHD star-disk numerical simulations \citep{ferr13a}.  

\begin{figure}
\centering
\includegraphics[width=1.0\textwidth]{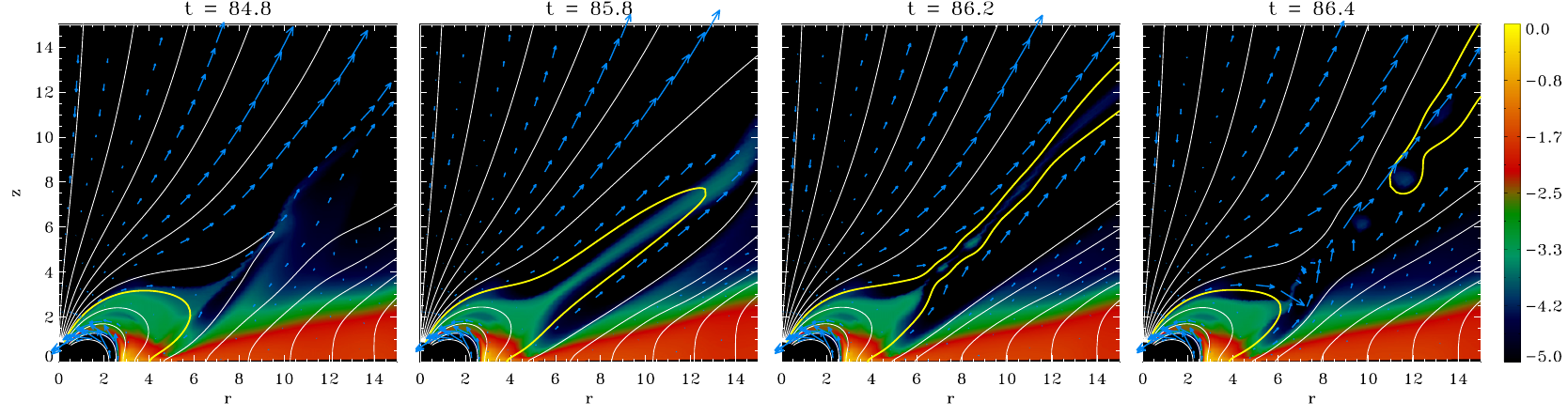}
\caption{Temporal evolution of the periodic dynamical processes leading to magnetospheric ejections. Time is given in units of the rotation period of the star, background colors are logarithmic density maps, white solid lines are sampled field lines, blue vectors show the poloidal speed. The yellow solid lines allow us to follow how the system evolves with time. As long as these field lines remain connected to both the star and the disk, angular momentum is being transferred from these two sources to plasma that is magneto-centrifugally ejected (as in a slingshot). This leads unavoidably to magnetic reconnection and plasmoid ejection, channelled between the inner stellar wind and the outer disk wind. Credit: \citet{zann13}.}
\label{Zanni+Ferreira13}
\end{figure}

2-D and 3-D numerical simulations of the star-disk interaction show that the disk truncation radius is always found to be smaller than the co-rotation radius when accretion takes place (e.g.\ \citealt{roma02,bess08,zann09,roma12} to cite only a few). However, some of these simulations do show the presence of ejecta from the disk truncation radius, although these events cannot be described within a steady-state MHD jet framework \citep{good97,roma09,zann13}. The latter authors showed that these magnetospheric ejections (referred to as MEs and see Fig.\ref{Zanni+Ferreira13}) have two major effects. First, they provide a strong spin down torque on both the star and the underlying disk, so that the combination of MEs and the inner stellar wind could indeed lead to the long awaited spin down torque on the star. Second, since these MEs are plasmoids resulting from reconnection episodes, they propagate almost ballistically and could not represent, alone, the YSO jet phenomenology. At best, they would be an additional, intermittent, plasma component flowing inside a narrow channel delimited between two quasi-steady MHD flows, namely the inner (on the axis) stellar wind and the outer disk wind.

\subsection{Disk winds and their link to accretion}

For quite a long time after the seminal paper of \citet{shak73}, the theory of accretion disk was only based on hydrodynamics and magnetic fields played no role. Plasmas in circumstellar accretion disk were assumed to behave like a neutral fluid, subjected to some turbulence that would (hopefully) provide the required anomalous viscosity. On the other hand, people working on astrophysical jets soon realized that self collimation can only be obtained if large scale magnetic fields are present, requiring the use of magnetohydrodynamics \citep{chan80,blan82}. This led \citet{pudr86} to suggest bipolar outflows are magnetically driven winds emitted from the surface of the molecular disk, providing a torque and leading to accretion. Thus, instead of being evacuated radially by some yet unknown turbulence, the disk angular momentum is transported vertically in the bipolar jets. 

\begin{figure}
\centering
$ \begin{array}{cc}
	 \includegraphics[width=0.47\textwidth]{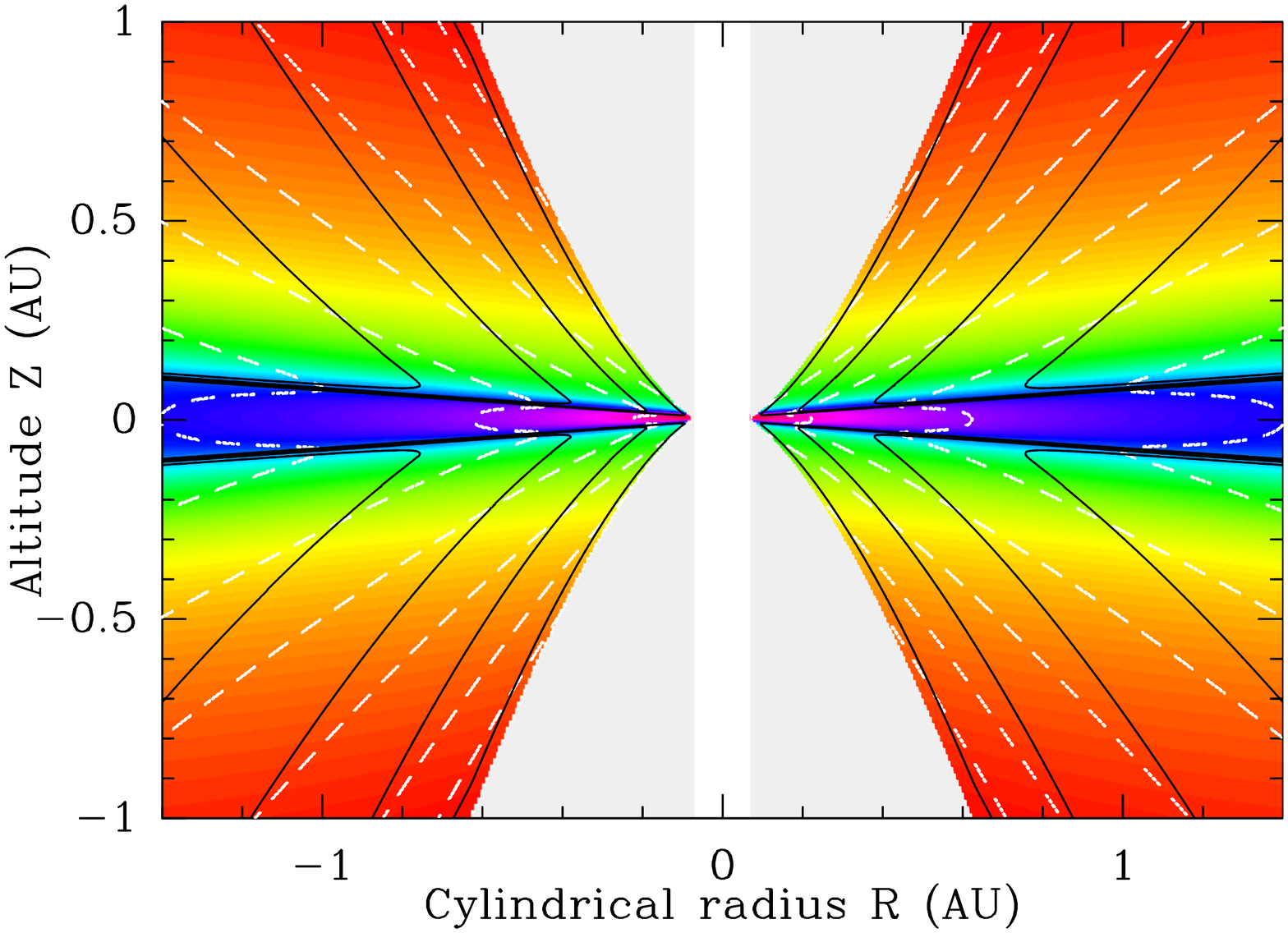} 
  	& 
  	\includegraphics[width=0.47\textwidth]{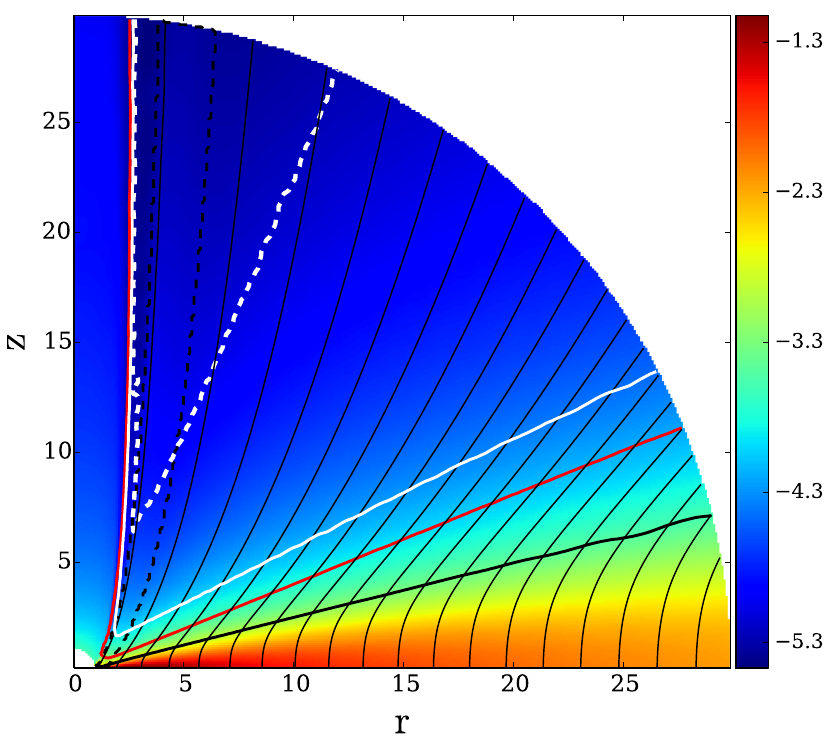}
\end{array}$
\caption{Close-up views on the disk of magnetized disk winds.  
{\bf Left}: Self-similar solution of an isothermal accretion-ejection structure \citep{ferr97}. Colours show the density, black solid lines show the streamlines  and dashed lines isocontours of the total velocity. In the resistive MHD disk (violet-blue region), matter is mostly accreting and rotating (rotation slows down vertically). Above the disk, plasma flows along the field lines (ideal MHD regime) and is progressively accelerated.    
{\bf Right}: Final output of a numerical simulation made by \citet{step16}. Colours show the density, thin black lines the poloidal magnetic field.  The lines mark the disk surface (black), the sonic surface (red) the Alfv\'en surface (white), and the fast surface (dashed white). It is striking to recover (away from the axis) conical critical surfaces in the final state, as in analytical self-similar studies, from a time dependent simulation. With kind permission of the authors.}
\label{Dwinds}
\end{figure}

The first tentative attempt to couple the \citet{blan82} MHD jet model to the outer regions of a circumstellar accretion disk was made by \citet{koni89}. Indeed, since jets exert a torque on the underlying disk, both its radial and vertical structures must be affected. Moreover, bipolar jets require the existence of a large scale vertical magnetic field threading the disk everywhere or only in the (probably) innermost region. This early work was then followed by many more efforts, of increasing complexity as regards disk-jet interrelations, \citep{ferr93a,ferr95,li95,ferr97,cass00b,ogil01,2019MNRAS.490.3112J} or in the inclusion of non-ideal MHD effects within the disk \citep{ward93,li96a,salm11}. 

In parallel to these analytical approaches, huge efforts have been made to provide numerical MHD simulations of jets emitted from accretion disks. The first attempts did not take the disk into account and computed the jet acceleration and propagation with some fixed boundary conditions at the base \citep{ouye97a,ouye97b,kras99,kras03,ouye03}. Such a "platform" approach is very promising for studying the jet collimation issue, allowing for instance to relate jet asymptotic properties to the radial distributions of jet mass loss and disk magnetic field that are imposed at the base \citep{ande05,pudr06,fend06}, up to observable scales \citep{staf10,rams11,staf15}. However, since the disk is not computed there is no feedback from the jet on the underlying disk structure, leaving thereby too much freedom on the imposed radial distributions (and no constraint on any possible inconsistency). The first MHD simulations that computed simultaneously the accretion disk and their jets are due to \citet{cass02,cass04}. Building upon the analytical results on the disk-jet interrelations found by \citet{ferr97}, they obtained super fast-magnetosonic jets using an $\alpha$ prescription for the turbulent disk. Their results have then been extended by other groups using different MHD codes \citep{zann07,tzef09,murp10,shei12,tzef13,step16}. Figure~\ref{Dwinds} provides an illustration of such findings. More recently 3-D MHD simulations of accretion and jet launching have been performed not only from single YSOs but also from disks in binary systems \citep{2015ApJ...814..113S, 2018ApJ...861...11S}. Here instabilities in the disk and jet are seen such as disk spiral arms and warps or jet precession. In turn this may lead to enhanced accretion rates. 

The main caveat of these $\alpha$-disk simulations is their possible inconsistency with MHD turbulence, since all anomalous transport coefficients (viscosity, magnetic diffusivity) are parametrized. Indeed, since the discovery of the Magneto-Rotational Instability or MRI by \citet{balb91}, it is known that magnetized accretion disks can sustain turbulence that acts effectively as a viscosity (see \citealt{hawl95}, \citealt{1996ApJ...463..656S}, \citealt{balb03} and references therein). It thus became clear that magnetic fields were also required in accretion disks, regardless of the jet phenomenon. MHD turbulence in a vertically stratified disk needs to be properly followed to make sure that simulations converge. Moreover, when a vertical magnetic field is included (instead of a toroidal one), mass loss is systematically observed from the disk surface and an outflow is obtained. As a consequence, the size of the computational domain needs to be large enough so that boundary conditions do not affect (or at least not too much) the outcome of the simulation. These are the main reasons why early shearing box simulations could hardly address MRI with a non zero net magnetic flux (see for instance \citealt{lesu13,from13,bai13} and references therein). Therefore, for quite a long time, the main focus of MRI studies was only the measurement of the Shakura-Sunayev $\alpha$ parameter describing the turbulent viscosity and the influence of non-ideal MHD effects. 

The presence of MRI relies on a small degree of ionization being present in the disk, so that in the outer cold regions the magnetic field is essentially coupled to the neutrals through ion-neutral collisions. It turns out that the MRI becomes quenched in the presence of non-ideal MHD effects, such as ambipolar diffusion (ion-neutral drift) and the Hall effect (ion-electron drift), both terms that are present whenever the plasma is not ionized enough \citep{gamm96,2013ApJ...769...76B,lesu14}. In addition expected sources of ionization, for example low and moderate energy galactic cosmic rays (E$_{\rm CR} \leq$ 1~GeV), cannot penetrate very far into a disk and may even be excluded by the strong magnetosphere of a young star \citep{2013ApJ...772....5C}.  That said high energy particles generated by the young star itself may at least assist ionization in the inner regions of the disk \citep{2017MNRAS.472...26R}. Nevertheless, in regions that are both too cold for collisional ionization and too dense for cosmic ray or radiative UV-X ionization, such as the disk mid-plane, no MRI can be triggered. As a consequence no $``$viscously$''$ driven accretion can take place, leading to the concept of "dead zones" \citep{gamm96}. Early simulations did not include non-ideal MHD effects, but when included these confirm the difficulty of providing viscosity through MRI generated turbulence \citep[e.g.][]{2015ApJ...798...84B}, or even directly using multi-fluid simulations \citep{okee14}. 

However, as already pointed out in \citet{gamm96}, MRI could remain active within thin layers close to the disk surface, thanks to cosmic rays ionization for instance. The existence of equatorial "dead zones" with upper and lower layers of MRI-driven accretion were thus thought to be characteristic of YSO disks, leading possibly to planet formation. This paradigm however proved unfruitful when it was realized that layered MRI-driven accretion could not properly account for the observed accretion rates in TTS \citep{bai11b}. Moreover, high spectral resolution observations, at least of the outer zones (tens of au) of disks with ALMA also show no evidence for the expected turbulence \citep{2018ApJ...856..117F}. Thus both simulations and observations suggest one should look for an alternative to MRI generated turbulence as the source of viscosity, and hence accretion, in the outer parts of accretion disks. 

One way out of this dilemma is to rely on an additional mechanism, such as magnetized winds, to redistribute angular momentum. Moreover, magnetised winds are naturally obtained when a large scale vertical field is present in the disk \citep{suzu09,from13,bai13}. Such a situation has been carefully avoided in early MRI numerical studies, because of the unfortunate tendency of these codes to crash. Global 3-D MHD simulations are thus best suited to address MRI with vertical B$_{\rm z}$ field, but achieving reliable simulations of this kind is a herculean task, that has only been tackled quite recently \citep{suzu14,gres15,beth17,zhu18,wang18}. 

A fair statement on the current consensus on YSO disks, is that a large scale vertical magnetic field is now invoked everywhere, not only to explain jet formation but also to explain accretion onto the central star \citep{bai11b,bai16,scep18}. In some sense, magnetized winds and jets are the main agent allowing accretion in YSO circumstellar disks and not, as thought for a long time, a mere epiphenomenon \citep[e.g.][]{2019FrASS...6...54P}. The presence of such a large scale field could also be responsible for features such as gaps and rings seen with ALMA \citep{suri18}.     

The lack of ionization beyond 1 au in YSO accretion disks has triggered a wealth of studies on the effect of non-ideal MHD terms on the development of the MRI and this is still a very active area of research. For instance, recent stratified models, that include Ohmic and ambipolar diffusion only, have shown that the disk could become essentially laminar (i.e.\ not turbulent) in the mid-plane, with a strong magnetically driven outflow launched at the disk surface (\citealt{bai13, gress15} and references therein). The effects of including the Hall effect (which may become dominant between 1 and 10 au) remain uncertain, with some discrepancies between single fluid \citep{lesu14,beth16} and multi-fluid MHD simulations \citep{okee14}. However, while relevant in outer regions of protostellar accretion disks, innermost disk regions as well as disks around active galactic nuclei and X-ray binaries are ionized enough to ensure non-ideal MHD effects vanish. Under these circumstances, one needs to rely on another source for diffusion, which can only be MHD turbulence itself. 

Although a large scale vertical magnetic field remains the unavoidable, albeit invisible, main ingredient for jet launching, thermal effects could still play a major role. As first proposed by \citet{cass00b}, UV-X illumination from the central regions could heat up the surface layers of the outer parts of the accretion disks, leading to strong magneto-thermal disk winds. Such a situation could be generic to accretion disks threaded by a large scale vertical field, in AGN \citep{fuku10a}, YSOs \citep{bai16,wang18} and X-ray binaries \citep{chak16a}. It is slightly ironic to realize that such a scenario has already been envisioned to explain the ubiquitous low velocity emission of forbidden lines in YSOs \citep{kwan95,kwan97}. The novelty is however the growing realization of the existence of a magnetic flux threading the whole accretion disk. Such a field could be a relic of the original infall, modified by the complex interplay between advection and diffusion (see \citealt{take14} and references therein).

\subsection{Are jets the prominent aspect of a multi-component structure?}

In essence then, the current consensus seems to favor magnetized disk winds as the source of YSO jets. Such jets would be generated from the innermost disk regions down to the disk truncation radius, where the disk magnetization is expected to be highest. This launching zone is expected to exhibit distinct features that could be observationally tested \citep{comb08}. There are however several aspects that cannot be easily explained by these winds: jet asymmetries \citep{podi11,meln09}, the range in jet time scales from months to hundred of years \citep{hart07a,agra11}), the presence of a hot inner flow \citep{edwa03, edwa06}, a possible steady component seen in X-rays \citep{gued05,skin11,boni11}. Moreover, disk winds are unable by nature to spin down the central star, which is also an absolute requirement. The global picture is therefore more complex than just magnetized disk winds.

A simple explanation could be that jets are only one component of a much broader, multi-component, MHD structure harboring:
\begin{enumerate}[i]
\item a warm stellar wind on the axis, emitted from the star and partly powered by the accretion shock;
\item surrounded by a radially stratified disk wind (going from fast inner jets to outer massive and slow magneto-thermal winds),
\item with time dependent magnetospheric ejections at the star-disk interface, braking down the star and producing internal jet shocks. 
\end{enumerate}

\section{Distinguishing Between Models for Jet Launching: The Hunt for Jet Rotation}\label{Rotation}

Of particular interest as a means of testing rival models for jet launching is the observation, or lack thereof, of jet or outflow rotation. As emphasised already in the previous section, if jets are launched from a disk initially as an MHD wind, then they must carry away angular momentum. Of course this is true even in the case of stellar winds, the crucial difference is that the degree of angular momentum transported is a function of the launch radius: winds launched at larger radii, normally carry away more angular momentum than those launched from smaller ones. This gives rise to an important difference between outflows/winds launched say at the co-rotation radius, e.g.\, in the X-wind model, in comparison to those generated at distances of several au in a disk. The latter will contain more angular momentum flux per unit mass. Of course for any reasonable density profile, a protoplanetary disk will contain more angular momentum per unit mass in its periphery than towards its inner radius. There is in fact a clear analogy here with the Solar System where the Sun contains the bulk of the mass but the planets contain most of the angular momentum. 

Observations of jet rotation can therefore give us clues as to where a jet was launched from. Early searches for rotation were made with the Space Telescope Imaging Spectrograph (STIS) on HST \citep{2002ApJ...576..222B, 2007ApJ...663..350C}. These were done as close as possible to the young star itself, in order to minimize any 
chances of a rotation signature being confused with, for example, the \say{buffeting} effects of the immediate environment. Moreover high spatial resolution was consider very important as essentially one is trying to distinguish differences in velocity laterally across a jet the width of which was at most a fraction of an arcsecond \citep{2000A&A...357L..61D}. 

The need to look close to the source implied classical T Tauri stars with jets, such as DG~Tau, were optimum targets. In a MHD disk wind model, for example, where the bulk of the jet is launched within approximately 1~au, the jet can achieve a poloidal velocity of a few hundred \kms . In such a scenario however we would then expect velocities differences, from one side of the jet to the other, of at most a few tens of \kms as a result of angular momentum transport. Although a number of claims of jet rotation have been made using HST, alternative explanations for the findings have been proposed including precession \citep{2006AIPC..875..285C} and asymmetrical jet shocks \citep{2016ApJ...832..152D}. Moreover the observations were in some cases confusing with, for example, results suggesting the jet was rotating in the opposite sense to the disk \citep[e.g.,][]{2016A&A...596A..88L}. Such findings suggested that either optical/near UV measurements are not an appropriate means of detecting rotation in jets or alternatively the assumption of a steady-state wind is not correct. 

\begin{figure}
\begin{center}
\includegraphics[width=12cm]{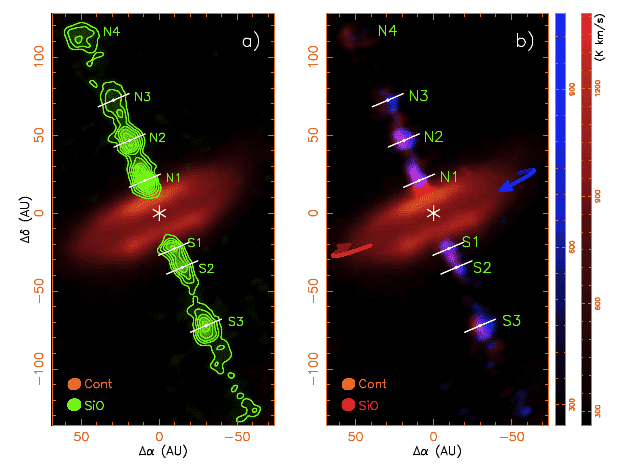}
\end{center}
\caption{ALMA observations of the launching region of the HH\,212 outflow. On the left we can see knots traced in SiO line emission while in red the continuum emission due to dust is observed. The latter traces the disk associated with the HH\,212 source. The sense of rotation of the disk is shown on the right and the same sense of rotation is seen in the knots to the north (N) and south (S) with blueshifted SiO emission to the right of each knot and redshifted SiO emission to the left. From \cite{2017NatAs...1E.152L} reproduced with permission \textcircled{c} Springer Nature.   } \label{HH212}
\end{figure}

Another means of achieving high spatial resolution, and even probing the outflow region close to the source where extinction might be large, is to use millimeter interferometry. Using this approach, observations of rotation have been claimed for example in the large-scale molecular outflow from Orion Source I with ALMA \citep{2017NatAs...1E.146H}. Perhaps the best example of rotation observed to date in a molecular outflow is in HH\,212 shown in Figure \ref{HH212}. Here ALMA detected velocity gradients across the outflow within approximately 100\,au from the embedded YSO. While the sense of rotation of all of the knots is the same, and mirrors that of the disk itself, the amount of rotation observed is small. In fact by assuming conservation of angular momentum along streamlines  \citep{2003ApJ...590L.107A} one can deduce that the outflow originates from within $\approx 0.05$\,au. Such values are consistent with either a close-in disk wind or even possibly an X-wind.

Evidence for jet rotation have been found in a number of other outflows. These include DG~Tau \citep{2011A&A...532A..59A}, , TMC\,1A6 \citep{2016Natur.540..406B} and HH30 \citep{2018A&A...618A.120L}. In a number of cases such observations suggest that the outflow originates on scales of a few au.

\section{Exploring the Launch Zone Using Optical/NIR Interferometers}
As already stated, we still do not know, with a high degree of certainty, if jets from young stars are launched in the form of a very strong stellar wind, from the zone where the magnetosphere of the star interacts with the circumstellar disk or from the disk itself. While we can get some clues from observing the region where the jet reaches its terminal velocity (i.e.\ at tens of au) and, in particular, from measuring its angular momentum content as explained in Section \ref{Rotation}, nevertheless imaging a jet on sub-au scales is clearly the best approach if one is to decisively determine how, and from where, they are launched. Moreover, it is not just spatial but velocity information that is required since different models predict very different kinematic behaviour in the vicinity of the launch zone. This type of data can only be obtained using optical/near-infrared spectro-interferometry, where not only the continuum from the disk but also line emission is spectrally and spatially resolved. 

Until recently a combination of technical challenges ensured that optical/NIR interferometers could only be used if the target is very bright  \citep[e.g.,][]{2012A&ARv..20...53B}. Major improvements in their design however have pushed the limiting magnitudes to the point where it is now possible not only to observe the immediate circumstellar environment of moderately bright Herbig Ae/Be stars (intermediate mass young stars) in the optical/near-infrared \citep{2015A&A...582A..44C} but also cTTS \citep{2017A&A...608A..78G}. 
Clearly to probe the jet on high resolution scales an appropriate diagnostic line, or better still a combination of lines, is required. Forbidden transitions, including those in the NIR, e.g., the [FeII] 1.64$\mu$m line, are unsuitable as their critical electron density is exceeded in an outflow when exploring regions very close to the source. In the past permitted HI lines (e.g.\ H$\alpha$, Pa$\beta$, and Br$\gamma$) were regarded solely as a measure of magnetospheric accretion \citep[e.g.,][and see Section \ref{Link-Accretion-Outflow}]{2004AJ....128.1294C, 2004A&A...424..603N}.  Nevertheless, we now know that such lines contain a mixture of accretion and outflow signatures \citep[e.g.,][]{2010A&A...522A.104L}, i.e.\ they are composite as demonstrated many years ago using spectro-astrometry \citep{2004A&A...417..247W}. Thus permitted lines can be used to produce rudimentary $''$images$''$ of the launch zone with an interferometer and may even $''$image$''$ the accretion zone if the interferometer has sufficiently high spatial resolution. 

In this way the new generation of spectro-interferometric instruments, such as GRAVITY and CHARA \citep{2011Msngr.143...16E, 2019A&A...627A..36L}, given their high spatial ($\approx$ a few milliarcseconds) and moderate spectral resolution ($R = \lambda/\delta\lambda$ up to a few thousand), as well as their higher sensitivity, are starting to permit us to measure the extent, position and velocity of the line emitting regions close to the source. In turn this allows us to start to discriminate between the various theoreis for their origin as the models (stellar, X- and disk-wind) predict the outflow emission to have different initial spatial widths and kinematic properties.

For example, if disk wind theory is correct, then one expects to see line emission, e.g.\ due to Br~$\gamma$, from an extended region (up to a few au) straddling the source in the disk plane, whereas if the X-wind model is closer to reality this emission will be much more confined (several tens of stellar radii). In both case very compact emission is also expected to coincide with the star`s magnetospheric accretion zone.  Moreover, for lines that are spectrally resolved, it should be possible to determine spatial offsets from the source as a function of velocity on sub-milliarcsecond scales. Potentially this could reveal the changing kinematics of the gas as it is launched from the star or disk. For example disk wind models predict a gradual change in velocity (both in magnitude and direction) as the wind leaves the disk surface at quasi-Keplerian speeds and then accelerates towards the Alfv{\'e}n radius \citep{2009ASSP...13...99F}. 

As mentioned above, optical/NIR interferometry is most easily attempted when the target is bright and so it is no surprise that Herbig Ae/Be stars where the first YSOs to be studied in this way \citep[e.g.,][]{2011A&A...527A.103W}. Interestingly extended, i.e.\ well beyond the expected co-rotation radius, Br$\gamma$ emission is found suggesting at least some of the flux comes from a disk wind \citep{2016MNRAS.456..156G} and similar results are found in the optical regime \citep{2016A&A...596A..17P}. While challenging in comparison to the  bright Herbig Ae/Be stars, studies with interferometers have been made of the much fainter T Tauri stars. Early attempts, using a single baseline, revealed only very compact line emission within the dust sublimation zone \citep{2014MNRAS.443.1916E}, and likely showed only the accretion component.  More recent studies however show that extended disk wind structures are present along with a magnetospheric accretion component \citep{2017A&A...608A..78G} and that the disk wind can be dusty \citep{2019A&A...627A..36L}. The wind component however is only seen at the earliest phases and once the T Tauri star becomes more evolved the accretion component is only observed \citep[][, Garcia-Lopez et al., 2020, in press]{2020A&A...636A.108B}.

\section{Future Observational Prospects}

Over the next decade there are a number of new instruments coming online that will no doubt make an impact on our understanding of the launching and propagation of outflows from young stars. It is not possible in this review to cover all of them in any depth. Instead we point out a few that we think might be important. 

We begin by mentioning the High Resolution Mid-InfrarEd Spectrometer (HIRMES) which has been developed by NASA Goddard for the Stratospheric Observatory for Infrared Astronomy (SOFIA). HIRMES is optimized to detect neutral atomic oxygen, water, as well as normal and deuterated hydrogen between 28 and 112$\mu$m \citep{2018JAI.....740015R}. A primarily scientific goal is to investigate the molecular and atomic gas content of protoplanetary disks as well as their ice features, e.g.\ at 43, 47 and 62$\mu$m. HIRMES, however, also offers the possibility of probing atomic jets launched from embedded systems using the [OI]\,63$\mu$m line previously observed, for example, using Herschel/PACS (see \ref{Case of HH211}). Of particular interest is the fact that 
HIRMES has a medium (R $\approx$ 12,000) spectral resolution mode making it ideal to study the kinematics of atomic jets from embedded sources in contrast to Herschel/PACS. It is hoped to commission the instrument in early 2021. 

At the time of writing, the launch of the James Webb Space Telescope (JWST), scheduled for March 2021, is eagerly awaited. A number of guaranteed time programs, and no doubt some Cycle~1 time, will be devoted to observing outflows from young stars as well as probing their circumstellar disks. Both JWST's high spatial resolution near-infrared camera and spectrograph, NIRCAM and NIRSPEC respectively, will be invaluable tools for such studies. Of particular importance will be the work undertaken using the Mid-Infrared Instrument (MIRI). MIRI has an integral field unit (IFU) spectrograph covering the range from approximately 5\,$\mu$m to 28\,$\mu$m and with a spectral resolution ($\lambda/\Delta\lambda$) of roughly 2,500. The combination of this and its high spatial resolution in the mid-infrared, should make it a very valuable instrument to probe not only conditions in YSO disks (within a few au from the central object) but also of outflows from the most embedded sources (e.g. jets from Class 0/I YSOs). Of note is the intention to map most of the blue lobe of the HH\,211 outflow (see \ref{Case of HH211}) with the MIRI IFU, the source itself, and some of its red lobe. This will require an extensive mosaic and almost 20\,hours of JWST time. Other embedded outflow sources will also be mapped but with much smaller mosaics and in addition a large amount of guaranteed time will be devoted to understanding disk evolution using MIRI. 

In order to advance models for wind/jet launching from disks, and even their parent stars, we need a handle on the magnetic field direction and strength in the launch zone. In the case of the least embedded sources, and normally this means weak-line T Tauri stars, we do have an idea of what their surface magnetic field values are like from optical spectro-polarimetry of both absorption and occasionally emission features in their spectrum \citep{2014IAUS..302...25H}. Here average surface fields in the several kG range are often seen and, somewhat surprisingly, the overall field structure is not necessarily bipolar as sometimes more complex geometries are seen. Unfortunately many of the sources studied to date are relatively evolved and lack an outflow with the exception of AA\,Tau \citep{2010MNRAS.409.1347D, 2013ApJ...762...40C} which possesses a faint jet discovered with HST. This situation however is about to change. A new spectro-polarimeter however has become available on the Canada France Hawaii Telescope (CFHT): known as SPIRou (Spectro-Polarimeter Infra-ROuge) it will utilize the near-infrared part of the spectrum to investigate not only more embedded, and hence less evolved, sources but also the inner zones of their associated disks. Since spectro-polarimetry exploits the Zeeman effect to measure magnetic fields, and the line separation grows with $\lambda^{\rm 2}$ all else being equal,  it is easier to measure magnetic field value in the NIR in comparison with the optical. Certainly a number of young stars with outflows are being targeted as part of the SPIRou Legacy Survey \citep{2018haex.bookE.107D}. 

In the past few years it is becoming increasingly obvious that radio observations can give us valuable information on jets from young stars and, in particular, their magnetic field geometry and even strength in those cases where the emission is non-thermal. As stated earlier on, in part this is because of the increased sensitivity of radio interferometers. Of course the next major advance at these wavelengths will come when the Square Kilometer Array (SKA) and next generation VLA (ngVLA) enter service. With these new facilities, we will be able to explore the jet launching zone with much better spatial resolution than before and this will be important from the point of view of distinguishing between various models for the collimation of jets within tens of au from the source. 

The higher sensitivity of the SKA and ngVLA may also allow us to hopefully employ a new diagnostic technique for imaging and kinematic purposes, i.e.\ using radio recombination lines (RRLs) instead of the broadband continuum that we now observe. RRL allow us not only to peer into the heart of very embedded, and hence otherwise inaccessible YSOs, but also to understand their dynamics \citep[see, for example,][] {2017ApJ...837...53Z}. 
At present only a few observations of RRL jet emission from massive young stars have been reported and these are almost certainly contaminated with radio emission from an underlying HII region \citep{2013ApJ...764L...4J, 2018ASPC..517..309G}. The future however is bright!

\section*{Funding}
TPR would like to acknowledge funding from the European Research Council under Advanced Grant No. 743029, Ejection, Accretion Structures in YSOs (EASY). 

\section*{Acknowledgments}
We thank the Editors for their invitation to write this review, and in particular Rob Fender for his patience and ongoing support of this work. We also wish to acknowledge useful comments from Christian Fendt, on an early version of this manuscript, and from an anonymous referee. 

\bibliographystyle{apalike} 
\bibliography{bibreview.bib}

\begin{thebibliography}{}

\bibitem[{Agra-Amboage} et~al., 2011a]{agra11}
{Agra-Amboage}, V., {Dougados}, C., {Cabrit}, S., and {Reunanen}, J. (2011a).
\newblock {Sub-arcsecond [Fe ii] spectro-imaging of the DG Tauri jet. Periodic
  bubbles and a dusty disk wind?}
\newblock {\em \aap}, 532:A59.

\bibitem[{Agra-Amboage} et~al., 2011b]{2011A&A...532A..59A}
{Agra-Amboage}, V., {Dougados}, C., {Cabrit}, S., and {Reunanen}, J. (2011b).
\newblock {Sub-arcsecond [Fe ii] spectro-imaging of the DG Tauri jet. Periodic
  bubbles and a dusty disk wind?}
\newblock {\em \aap}, 532:A59.

\bibitem[{Ainsworth} et~al., 2014]{2014ApJ...792L..18A}
{Ainsworth}, R.~E., {Scaife}, A. M.~M., {Ray}, T.~P., {Taylor}, A.~M., {Green},
  D.~A., and {Buckle}, J.~V. (2014).
\newblock {Tentative Evidence for Relativistic Electrons Generated by the Jet
  of the Young Sun-like Star DG Tau}.
\newblock {\em \apjl}, 792(1):L18.

\bibitem[{Alcal{\'a}} et~al., 2017]{2017A&A...600A..20A}
{Alcal{\'a}}, J.~M., {Manara}, C.~F., {Natta}, A., {Frasca}, A., {Testi}, L.,
  {Nisini}, B., {Stelzer}, B., {Williams}, J.~P., {Antoniucci}, S., {Biazzo},
  K., {Covino}, E., {Esposito}, M., {Getman}, F., and {Rigliaco}, E. (2017).
\newblock {X-shooter spectroscopy of young stellar objects in Lupus. Accretion
  properties of class II and transitional objects}.
\newblock {\em \aap}, 600:A20.

\bibitem[{Anderson} et~al., 2003]{2003ApJ...590L.107A}
{Anderson}, J.~M., {Li}, Z.-Y., {Krasnopolsky}, R., and {Blandford}, R.~D.
  (2003).
\newblock {Locating the Launching Region of T Tauri Winds: The Case of DG
  Tauri}.
\newblock {\em \apjl}, 590(2):L107--L110.

\bibitem[{Anderson} et~al., 2005]{ande05}
{Anderson}, J.~M., {Li}, Z.-Y., {Krasnopolsky}, R., and {Blandford}, R.~D.
  (2005).
\newblock {The Structure of Magnetocentrifugal Winds. I. Steady Mass Loading}.
\newblock {\em \apj}, 630:945--957.

\bibitem[{Andre} et~al., 1993]{1993ApJ...406..122A}
{Andre}, P., {Ward-Thompson}, D., and {Barsony}, M. (1993).
\newblock {Submillimeter Continuum Observations of rho Ophiuchi A: The
  Candidate Protostar VLA 1623 and Prestellar Clumps}.
\newblock {\em \apj}, 406:122.

\bibitem[{Anglada} et~al., 2018]{2018A&ARv..26....3A}
{Anglada}, G., {Rodr{\'{\i}}guez}, L.~F., and {Carrasco-Gonz{\'a}lez}, C.
  (2018).
\newblock {Radio jets from young stellar objects}.
\newblock {\em \aapr}, 26:3.

\bibitem[{Antoniucci} et~al., 2014]{2014A&A...572A..62A}
{Antoniucci}, S., {Garc{\'\i}a L{\'o}pez}, R., {Nisini}, B., {Caratti o
  Garatti}, A., {Giannini}, T., and {Lorenzetti}, D. (2014).
\newblock {POISSON project. III. Investigating the evolution of the mass
  accretion rate}.
\newblock {\em \aap}, 572:A62.

\bibitem[{Bacciotti} and {Eisl{\"o}ffel}, 1999]{1999A&A...342..717B}
{Bacciotti}, F. and {Eisl{\"o}ffel}, J. (1999).
\newblock {Ionization and density along the beams of Herbig-Haro jets}.
\newblock {\em \aap}, 342:717--735.

\bibitem[{Bacciotti} et~al., 2002]{2002ApJ...576..222B}
{Bacciotti}, F., {Ray}, T.~P., {Mundt}, R., {Eisl{\"o}ffel}, J., and {Solf}, J.
  (2002).
\newblock {Hubble Space Telescope/STIS Spectroscopy of the Optical Outflow from
  DG Tauri: Indications for Rotation in the Initial Jet Channel}.
\newblock {\em \apj}, 576(1):222--231.

\bibitem[{Bai}, 2011]{bai11b}
{Bai}, X.-N. (2011).
\newblock {Magnetorotational-instability-driven Accretion in Protoplanetary
  Disks}.
\newblock {\em \apj}, 739:50.

\bibitem[{Bai}, 2015]{2015ApJ...798...84B}
{Bai}, X.-N. (2015).
\newblock {Hall Effect Controlled Gas Dynamics in Protoplanetary Disks. II.
  Full 3D Simulations toward the Outer Disk}.
\newblock {\em \apj}, 798(2):84.

\bibitem[{Bai} and {Stone}, 2013a]{bai13}
{Bai}, X.-N. and {Stone}, J.~M. (2013a).
\newblock {Local Study of Accretion Disks with a Strong Vertical Magnetic
  Field: Magnetorotational Instability and Disk Outflow}.
\newblock {\em \apj}, 767:30.

\bibitem[{Bai} and {Stone}, 2013b]{2013ApJ...769...76B}
{Bai}, X.-N. and {Stone}, J.~M. (2013b).
\newblock {Wind-driven Accretion in Protoplanetary Disks. I. Suppression of the
  Magnetorotational Instability and Launching of the Magnetocentrifugal Wind}.
\newblock {\em \apj}, 769(1):76.

\bibitem[{Bai} et~al., 2016]{bai16}
{Bai}, X.-N., {Ye}, J., {Goodman}, J., and {Yuan}, F. (2016).
\newblock {Magneto-thermal Disk Winds from Protoplanetary Disks}.
\newblock {\em \apj}, 818:152.

\bibitem[{Balbus}, 2003]{balb03}
{Balbus}, S.~A. (2003).
\newblock {Enhanced Angular Momentum Transport in Accretion Disks}.
\newblock {\em \araa}, 41:555--597.

\bibitem[{Balbus} and {Hawley}, 1991]{balb91}
{Balbus}, S.~A. and {Hawley}, J.~F. (1991).
\newblock {A powerful local shear instability in weakly magnetized disks. I -
  Linear analysis. II - Nonlinear evolution}.
\newblock {\em \apj}, 376:214--233.

\bibitem[{Bally} et~al., 2012]{2012AJ....144..143B}
{Bally}, J., {Walawender}, J., and {Reipurth}, B. (2012).
\newblock {Deep Imaging Surveys of Star-forming Clouds. V. New Herbig-Haro
  Shocks and Giant Outflows in Taurus}.
\newblock {\em \aj}, 144(5):143.

\bibitem[{Berger} et~al., 2012]{2012A&ARv..20...53B}
{Berger}, J.~P., {Malbet}, F., {Baron}, F., {Chiavassa}, A., {Duvert}, G.,
  {Elitzur}, M., {Freytag}, B., {Gueth}, F., {H{\"o}nig}, S., {Hron}, J.,
  {Jang-Condell}, H., {Le Bouquin}, J.~B., {Monin}, J.~L., {Monnier}, J.~D.,
  {Perrin}, G., {Plez}, B., {Ratzka}, T., {Renard}, S., {Stefl}, S.,
  {Thi{\'e}baut}, E., {Tristram}, K.~R.~W., {Verhoelst}, T., {Wolf}, S., and
  {Young}, J. (2012).
\newblock {Imaging the heart of astrophysical objects with optical
  long-baseline interferometry}.
\newblock {\em \aapr}, 20:53.

\bibitem[{Bertout}, 1989]{bert89}
{Bertout}, C. (1989).
\newblock {T Tauri stars: wild as dust.}
\newblock {\em \araa}, 27:351--395.

\bibitem[{Bessolaz} et~al., 2008]{bess08}
{Bessolaz}, N., {Zanni}, C., {Ferreira}, J., {Keppens}, R., and {Bouvier}, J.
  (2008).
\newblock {Accretion funnels onto weakly magnetized young stars}.
\newblock {\em \aap}, 478:155--162.

\bibitem[{B{\'e}thune} et~al., 2016]{beth16}
{B{\'e}thune}, W., {Lesur}, G., and {Ferreira}, J. (2016).
\newblock {Self-organisation in protoplanetary discs. Global, non-stratified
  Hall-MHD simulations}.
\newblock {\em \aap}, 589:A87.

\bibitem[{B{\'e}thune} et~al., 2017]{beth17}
{B{\'e}thune}, W., {Lesur}, G., and {Ferreira}, J. (2017).
\newblock {Global simulations of protoplanetary disks with net magnetic flux.
  I. Non-ideal MHD case}.
\newblock {\em \aap}, 600:A75.

\bibitem[{Bjerkeli} et~al., 2016]{2016Natur.540..406B}
{Bjerkeli}, P., {van der Wiel}, M.~H.~D., {Harsono}, D., {Ramsey}, J.~P., and
  {J{\o}rgensen}, J.~K. (2016).
\newblock {Resolved images of a protostellar outflow driven by an extended disk
  wind}.
\newblock {\em \nat}, 540:406--409.

\bibitem[{Blandford} and {Payne}, 1982]{blan82}
{Blandford}, R.~D. and {Payne}, D.~G. (1982).
\newblock {Hydromagnetic flows from accretion discs and the production of radio
  jets}.
\newblock {\em \mnras}, 199:883--903.

\bibitem[{Blandford} and {Rees}, 1974]{blan74}
{Blandford}, R.~D. and {Rees}, M.~J. (1974).
\newblock {A ``twin-exhaust'' model for double radio sources.}
\newblock {\em \mnras}, 169:395--415.

\bibitem[{Bonito} et~al., 2011]{boni11}
{Bonito}, R., {Orlando}, S., {Miceli}, M., {Peres}, G., {Micela}, G., and
  {Favata}, F. (2011).
\newblock {X-Ray Emission from Protostellar Jet HH 154: The First Evidence of a
  Diamond Shock?}
\newblock {\em \apj}, 737:54.

\bibitem[{Bosch-Ramon} et~al., 2010]{2010A&A...511A...8B}
{Bosch-Ramon}, V., {Romero}, G.~E., {Araudo}, A.~T., and {Paredes}, J.~M.
  (2010).
\newblock {Massive protostars as gamma-ray sources}.
\newblock {\em \aap}, 511:A8.

\bibitem[{Bouvier} et~al., 1986]{bouv86}
{Bouvier}, J., {Bertout}, C., {Benz}, W., and {Mayor}, M. (1986).
\newblock {Rotation in T Tauri stars. I - Observations and immediate analysis}.
\newblock {\em \aap}, 165:110--119.

\bibitem[{Bouvier} et~al., 2020]{2020A&A...636A.108B}
{Bouvier}, J., {Perraut}, K., {Le Bouquin}, J.~B., {Duvert}, G., {Dougados},
  C., {Brandner}, W., {Benisty}, M., {Berger}, J.~P., and {Al{\'e}cian}, E.
  (2020).
\newblock {Probing the magnetospheric accretion region of the young
  pre-transitional disk system DoAr 44 using VLTI/GRAVITY}.
\newblock {\em \aap}, 636:A108.

\bibitem[{Buehrke} et~al., 1988]{1988A&A...200...99B}
{Buehrke}, T., {Mundt}, R., and {Ray}, T.~P. (1988).
\newblock {A detailed study of HH 34 and its associated jet.}
\newblock {\em \aap}, 200:99--119.

\bibitem[{Cabrit}, 2007]{cabr07}
{Cabrit}, S. (2007).
\newblock {Jets from Young Stars: The Need for MHD Collimation and Acceleration
  Processes}.
\newblock In {Ferreira}, J., {Dougados}, C., and {Whelan}, E., editors, {\em
  Jets from Young Stars: Models and Constraints}, volume 723 of {\em Lecture
  Notes in Physics, Berlin Springer Verlag}, page~21.

\bibitem[{Cai} et~al., 2008]{cai08}
{Cai}, M.~J., {Shang}, H., {Lin}, H.-H., and {Shu}, F.~H. (2008).
\newblock {X-Winds in Action}.
\newblock {\em \apj}, 672:489--503.

\bibitem[{Calvet} and {Gullbring}, 1998]{1998ApJ...509..802C}
{Calvet}, N. and {Gullbring}, E. (1998).
\newblock {The Structure and Emission of the Accretion Shock in T Tauri Stars}.
\newblock {\em \apj}, 509(2):802--818.

\bibitem[{Calvet} et~al., 2004]{2004AJ....128.1294C}
{Calvet}, N., {Muzerolle}, J., {Brice{\~n}o}, C., {Hern{\'a}ndez}, J.,
  {Hartmann}, L., {Saucedo}, J.~L., and {Gordon}, K.~D. (2004).
\newblock {The Mass Accretion Rates of Intermediate-Mass T Tauri Stars}.
\newblock {\em \aj}, 128(3):1294--1318.

\bibitem[{Canto}, 1980]{cant80}
{Canto}, J. (1980).
\newblock {A stellar wind model for Herbig-Haro objects}.
\newblock {\em \aap}, 86(3):327--338.

\bibitem[{Caratti o Garatti} et~al., 2015]{2015A&A...582A..44C}
{Caratti o Garatti}, A., {Tambovtseva}, L.~V., {Garcia Lopez}, R., {Kraus}, S.,
  {Schertl}, D., {Grinin}, V.~P., {Weigelt}, G., {Hofmann}, K.~H., {Massi}, F.,
  {Lagarde}, S., {Vannier}, M., and {Malbet}, F. (2015).
\newblock {AMBER/VLTI high spectral resolution observations of the
  Br{\ensuremath{\gamma}} emitting region in HD 98922. A compact disc wind
  launched from the inner disc region}.
\newblock {\em \aap}, 582:A44.

\bibitem[{Carrasco-Gonz{\'a}lez} et~al., 2010]{2010Sci...330.1209C}
{Carrasco-Gonz{\'a}lez}, C., {Rodr{\'{\i}}guez}, L.~F., {Anglada}, G.,
  {Mart{\'{\i}}}, J., {Torrelles}, J.~M., and {Osorio}, M. (2010).
\newblock {A Magnetized Jet from a Massive Protostar}.
\newblock {\em Science}, 330:1209.

\bibitem[{Casse} and {Ferreira}, 2000]{cass00b}
{Casse}, F. and {Ferreira}, J. (2000).
\newblock {Magnetized accretion-ejection structures. V. Effects of entropy
  generation inside the disc}.
\newblock {\em \aap}, 361:1178--1190.

\bibitem[{Casse} and {Keppens}, 2002]{cass02}
{Casse}, F. and {Keppens}, R. (2002).
\newblock {Magnetized Accretion-Ejection Structures: 2.5-dimensional
  Magnetohydrodynamic Simulations of Continuous Ideal Jet Launching from
  Resistive Accretion Disks}.
\newblock {\em \apj}, 581:988--1001.

\bibitem[{Casse} and {Keppens}, 2004]{cass04}
{Casse}, F. and {Keppens}, R. (2004).
\newblock {Radiatively Inefficient Magnetohydrodynamic Accretion-Ejection
  Structures}.
\newblock {\em \apj}, 601:90--103.

\bibitem[{Cerqueira} et~al., 2006]{2006AIPC..875..285C}
{Cerqueira}, A.~H., {Vel{\'a}zquez}, P.~F., {Raga}, A.~C., {Vasconcelos},
  M.~J., and {de Colle}, F. (2006).
\newblock {Radial velocity asymmetries from jets with variable velocity
  profiles}.
\newblock In {Herrera-Vel{\'a}zquez}, J. J.~E., editor, {\em Plasma and Fusion
  Science: 16th IAEA Technical Meeting on Research using Small Fusion Devices},
  volume 875 of {\em American Institute of Physics Conference Series}, pages
  285--288.

\bibitem[{Chakravorty} et~al., 2016]{chak16a}
{Chakravorty}, S., {Petrucci}, P.-O., {Ferreira}, J., {Henri}, G., {Belmont},
  R., {Clavel}, M., {Corbel}, S., {Rodriguez}, J., {Coriat}, M., {Drappeau},
  S., and {Malzac}, J. (2016).
\newblock {Absorption lines from magnetically driven winds in X-ray binaries}.
\newblock {\em \aap}, 589:A119.

\bibitem[{Chan} and {Henriksen}, 1980]{chan80}
{Chan}, K.~L. and {Henriksen}, R.~N. (1980).
\newblock {On the supersonic dynamics of magnetized jets of thermal gas in
  radio galaxies}.
\newblock {\em \apj}, 241:534--551.

\bibitem[{Cleeves} et~al., 2013]{2013ApJ...772....5C}
{Cleeves}, L.~I., {Adams}, F.~C., and {Bergin}, E.~A. (2013).
\newblock {Exclusion of Cosmic Rays in Protoplanetary Disks: Stellar and
  Magnetic Effects}.
\newblock {\em \apj}, 772(1):5.

\bibitem[{Codella} et~al., 2007]{2007A&A...462L..53C}
{Codella}, C., {Cabrit}, S., {Gueth}, F., {Cesaroni}, R., {Bacciotti}, F.,
  {Lefloch}, B., and {McCaughrean}, M.~J. (2007).
\newblock {A highly-collimated SiO jet in the HH212 protostellar outflow}.
\newblock {\em \aap}, 462(3):L53--L56.

\bibitem[{Coffey} et~al., 2007]{2007ApJ...663..350C}
{Coffey}, D., {Bacciotti}, F., {Ray}, T.~P., {Eisl{\"o}ffel}, J., and {Woitas},
  J. (2007).
\newblock {Further Indications of Jet Rotation in New Ultraviolet and Optical
  Hubble Space Telescope STIS Spectra}.
\newblock {\em \apj}, 663(1):350--364.

\bibitem[{Combet} and {Ferreira}, 2008]{comb08}
{Combet}, C. and {Ferreira}, J. (2008).
\newblock {The radial structure of protostellar accretion disks: influence of
  jets}.
\newblock {\em \aap}, 479:481--491.

\bibitem[{Coughlan} et~al., 2017]{2017ApJ...834..206C}
{Coughlan}, C.~P., {Ainsworth}, R.~E., {Eisl{\"o}ffel}, J., {Hoeft}, M.,
  {Drabent}, A., {Scaife}, A. M.~M., {Ray}, T.~P., {Bell}, M.~E., {Broderick},
  J.~W., {Corbel}, S., {Grie{\ss}meier}, J.-M., {van der Horst}, A.~J., {van
  Leeuwen}, J., {Markoff}, S., {Pietka}, M., {Stewart}, A.~J., {Wijers}, R.
  A.~M.~J., and {Zarka}, P. (2017).
\newblock {A LOFAR Detection of the Low-mass Young Star T Tau at 149 MHz}.
\newblock {\em \apj}, 834(2):206.

\bibitem[{Cox} et~al., 2013]{2013ApJ...762...40C}
{Cox}, A.~W., {Grady}, C.~A., {Hammel}, H.~B., {Hornbeck}, J., {Russell},
  R.~W., {Sitko}, M.~L., and {Woodgate}, B.~E. (2013).
\newblock {Imaging the Disk and Jet of the Classical T Tauri Star AA Tau}.
\newblock {\em \apj}, 762:40.

\bibitem[{De Colle} et~al., 2016]{2016ApJ...832..152D}
{De Colle}, F., {Cerqueira}, A.~H., and {Riera}, A. (2016).
\newblock {Transverse Velocity Shifts in Protostellar Jets: Rotation or
  Velocity Asymmetries?}
\newblock {\em \apj}, 832(2):152.

\bibitem[{DeCampli}, 1981]{deca81}
{DeCampli}, W.~M. (1981).
\newblock {T Tauri winds}.
\newblock {\em \apj}, 244:124--146.

\bibitem[{Dionatos} et~al., 2018]{2018A&A...616A..84D}
{Dionatos}, O., {Ray}, T., and {G{\"u}del}, M. (2018).
\newblock {Herschel spectral-line mapping of the HH211 protostellar system}.
\newblock {\em \aap}, 616:A84.

\bibitem[{Donati} et~al., 2018]{2018haex.bookE.107D}
{Donati}, J.-F., {Kouach}, D., {Lacombe}, M., {Baratchart}, S., {Doyon}, R.,
  {Delfosse}, X., {Artigau}, {\'E}., {Moutou}, C., {H{\'e}brard}, G., {Bouchy},
  F., {Bouvier}, J., {Alencar}, S., {Saddlemyer}, L., {Par{\`e}s}, L., {Rabou},
  P., {Micheau}, Y., {Dolon}, F., {Barrick}, G., {Hernandez}, O., {Wang},
  S.~Y., {Reshetov}, V., {Striebig}, N., {Challita}, Z., {Carmona}, A.,
  {Tibault}, S., {Martioli}, E., {Figueira}, P., {Boisse}, I., and {Pepe}, F.
  (2018).
\newblock {\em Handbook of Exoplanets}, chapter {SPIRou: A NIR
  Spectropolarimeter/High-Precision Velocimeter for the CFHT}, page 107.
\newblock Springer International Publishing.

\bibitem[{Donati} et~al., 2010]{2010MNRAS.409.1347D}
{Donati}, J.-F., {Skelly}, M.~B., {Bouvier}, J., {Gregory}, S.~G., {Grankin},
  K.~N., {Jardine}, M.~M., {Hussain}, G.~A.~J., {M{\'e}nard}, F., {Dougados},
  C., {Unruh}, Y., {Mohanty}, S., {Auri{\`e}re}, M., {Morin}, J., {Far{\`e}s},
  R., and {MAPP Collaboration} (2010).
\newblock {Magnetospheric accretion and spin-down of the prototypical classical
  T Tauri star AA Tau}.
\newblock {\em \mnras}, 409:1347--1361.

\bibitem[{Dougados} et~al., 2000]{2000A&A...357L..61D}
{Dougados}, C., {Cabrit}, S., {Lavalley}, C., and {M{\'e}nard}, F. (2000).
\newblock {T Tauri stars microjets resolved by adaptive optics}.
\newblock {\em \aap}, 357:L61--L64.

\bibitem[{Edwards}, 1997]{1997IAUS..182..433E}
{Edwards}, S. (1997).
\newblock {Magnetospherically Mediated Accretion in Classical T Tauri Stars.
  Paradigm for Low Mass Stars Undergoing Disk Accretion?}
\newblock In {Reipurth}, B. and {Bertout}, C., editors, {\em Herbig-Haro Flows
  and the Birth of Stars}, volume 182 of {\em IAU Symposium}, pages 433--442.

\bibitem[{Edwards} et~al., 2006]{edwa06}
{Edwards}, S., {Fischer}, W., {Hillenbrand}, L., and {Kwan}, J. (2006).
\newblock {Probing T Tauri Accretion and Outflow with 1 Micron Spectroscopy}.
\newblock {\em \apj}, 646:319--341.

\bibitem[{Edwards} et~al., 2003]{edwa03}
{Edwards}, S., {Fischer}, W., {Kwan}, J., {Hillenbrand}, L., and {Dupree},
  A.~K. (2003).
\newblock {He I {$\lambda$}10830 as a Probe of Winds in Accreting Young Stars}.
\newblock {\em \apjl}, 599:L41--L44.

\bibitem[{Eisenhauer} et~al., 2011]{2011Msngr.143...16E}
{Eisenhauer}, F., {Perrin}, G., {Brandner}, W., {Straubmeier}, C., {Perraut},
  K., {Amorim}, A., {Sch{\"o}ller}, M., {Gillessen}, S., {Kervella}, P.,
  {Benisty}, M., {Araujo-Hauck}, C., {Jocou}, L., {Lima}, J., {Jakob}, G.,
  {Haug}, M., {Cl{\'e}net}, Y., {Henning}, T., {Eckart}, A., {Berger}, J.~P.,
  {Garcia}, P., {Abuter}, R., {Kellner}, S., {Paumard}, T., {Hippler}, S.,
  {Fischer}, S., {Moulin}, T., {Villate}, J., {Avila}, G., {Gr{\"a}ter}, A.,
  {Lacour}, S., {Huber}, A., {Wiest}, M., {Nolot}, A., {Carvas}, P., {Dorn},
  R., {Pfuhl}, O., {Gendron}, E., {Kendrew}, S., {Yazici}, S., {Anton}, S.,
  {Jung}, Y., {Thiel}, M., {Choquet}, {\'E}., {Klein}, R., {Teixeira}, P.,
  {Gitton}, P., {Moch}, D., {Vincent}, F., {Kudryavtseva}, N., {Str{\"o}bele},
  S., {Sturm}, S., {F{\'e}dou}, P., {Lenzen}, R., {Jolley}, P., {Kister}, C.,
  {Lapeyr{\`e}re}, V., {Naranjo}, V., {Lucuix}, C., {Hofmann}, R., {Chapron},
  F., {Neumann}, U., {Mehrgan}, L., {Hans}, O., {Rousset}, G., {Ramos}, J.,
  {Suarez}, M., {Lederer}, R., {Reess}, J.~M., {Rohloff}, R.~R., {Haguenauer},
  P., {Bartko}, H., {Sevin}, A., {Wagner}, K., {Lizon}, J.~L., {Rabien}, S.,
  {Collin}, C., {Finger}, G., {Davies}, R., {Rouan}, D., {Wittkowski}, M.,
  {Dodds-Eden}, K., {Ziegler}, D., {Cassaing}, F., {Bonnet}, H., {Casali}, M.,
  {Genzel}, R., and {Lena}, P. (2011).
\newblock {GRAVITY: Observing the Universe in Motion}.
\newblock {\em The Messenger}, 143:16--24.

\bibitem[{Eisloeffel} and {Mundt}, 1992]{1992A&A...263..292E}
{Eisloeffel}, J. and {Mundt}, R. (1992).
\newblock {Proper motion measurements in the HH 34 jet and its associated bow
  shocks.}
\newblock {\em \aap}, 263:292--300.

\bibitem[{Eisner} et~al., 2014]{2014MNRAS.443.1916E}
{Eisner}, J.~A., {Hillenbrand}, L.~A., and {Stone}, J.~M. (2014).
\newblock {Constraining the sub-au-scale distribution of hydrogen and carbon
  monoxide gas around young stars with the Keck Interferometer}.
\newblock {\em \mnras}, 443(3):1916--1945.

\bibitem[{Feeney-Johansson} et~al., 2019]{2019ApJ...885L...7F}
{Feeney-Johansson}, A., {Purser}, S. J.~D., {Ray}, T.~P., {Eisl{\"o}ffel}, J.,
  {Hoeft}, M., {Drabent}, A., and {Ainsworth}, R.~E. (2019).
\newblock {The First Detection of a Low-frequency Turnover in Nonthermal
  Emission from the Jet of a Young Star}.
\newblock {\em \apjl}, 885(1):L7.

\bibitem[{Fendt}, 2006]{fend06}
{Fendt}, C. (2006).
\newblock {Collimation of Astrophysical Jets: The Role of the Accretion Disk
  Magnetic Field Distribution}.
\newblock {\em \apj}, 651:272--287.

\bibitem[{Ferreira}, 1997]{ferr97}
{Ferreira}, J. (1997).
\newblock Magnetically-driven jets from keplerian accretion discs.
\newblock {\em \aap}, 319:340--359.

\bibitem[{Ferreira}, 2009]{2009ASSP...13...99F}
{Ferreira}, J. (2009).
\newblock {Self-Collimated Jets from Accretion Discs and Star-disc Interaction
  Zones}.
\newblock {\em Astrophysics and Space Science Proceedings}, 13:99--110.

\bibitem[{Ferreira} and {Casse}, 2013]{ferr13a}
{Ferreira}, J. and {Casse}, F. (2013).
\newblock {On fan-shaped cold MHD winds from Keplerian accretion discs}.
\newblock {\em \mnras}, 428:307--320.

\bibitem[{Ferreira} et~al., 2006]{ferr06b}
{Ferreira}, J., {Dougados}, C., and {Cabrit}, S. (2006).
\newblock {Which jet launching mechanism(s) for T Tauri stars?}
\newblock {\em \aap}, 453:785--796.

\bibitem[{Ferreira} and {Pelletier}, 1993]{ferr93a}
{Ferreira}, J. and {Pelletier}, G. (1993).
\newblock Magnetized accretion-ejection structures. 1. general statements.
\newblock {\em \aap}, 276:625.

\bibitem[{Ferreira} and {Pelletier}, 1995]{ferr95}
{Ferreira}, J. and {Pelletier}, G. (1995).
\newblock Magnetized accretion-ejection structures. iii. stellar and
  extragalactic jets as weakly dissipative disk outflows.
\newblock {\em \aap}, 295:807.

\bibitem[{Flaherty} et~al., 2018]{2018ApJ...856..117F}
{Flaherty}, K.~M., {Hughes}, A.~M., {Teague}, R., {Simon}, J.~B., {Andrews},
  S.~M., and {Wilner}, D.~J. (2018).
\newblock {Turbulence in the TW Hya Disk}.
\newblock {\em \apj}, 856(2):117.

\bibitem[{Fromang} et~al., 2013]{from13}
{Fromang}, S., {Latter}, H., {Lesur}, G., and {Ogilvie}, G.~I. (2013).
\newblock {Local outflows from turbulent accretion disks}.
\newblock {\em \aap}, 552:A71.

\bibitem[{Fukumura} et~al., 2010]{fuku10a}
{Fukumura}, K., {Kazanas}, D., {Contopoulos}, I., and {Behar}, E. (2010).
\newblock {Magnetohydrodynamic Accretion Disk Winds as X-ray Absorbers in
  Active Galactic Nuclei}.
\newblock {\em \apj}, 715:636--650.

\bibitem[{Gallet} and {Bouvier}, 2013]{gall13}
{Gallet}, F. and {Bouvier}, J. (2013).
\newblock {Improved angular momentum evolution model for solar-like stars}.
\newblock {\em \aap}, 556:A36.

\bibitem[{Galv{\'a}n-Madrid} et~al., 2018]{2018ASPC..517..309G}
{Galv{\'a}n-Madrid}, R., {Beltr{\'a}n}, M., {Ginsburg}, A.,
  {Carrasco-Gonz{\'a}lez}, C., {Liu}, H.~B., {Rodr{\'\i}guez}, L.~F., and
  {Kurtz}, S. (2018).
\newblock {Resolving the Structure and Kinematics of the Youngest HII Regions
  and Radio Jets from Young Stellar Objects}.
\newblock In {Murphy}, E., editor, {\em Science with a Next Generation Very
  Large Array}, volume 517 of {\em Astronomical Society of the Pacific
  Conference Series}, page 309.

\bibitem[{Gammie}, 1996]{gamm96}
{Gammie}, C.~F. (1996).
\newblock {Layered Accretion in T Tauri Disks}.
\newblock {\em \apj}, 457:355.

\bibitem[{Garcia Lopez} et~al., 2016]{2016MNRAS.456..156G}
{Garcia Lopez}, R., {Kurosawa}, R., {Caratti o Garatti}, A., {Kreplin}, A.,
  {Weigelt}, G., {Tambovtseva}, L.~V., {Grinin}, V.~P., and {Ray}, T.~P.
  (2016).
\newblock {Investigating the origin and spectroscopic variability of the
  near-infrared H I lines in the Herbig star VV Ser}.
\newblock {\em \mnras}, 456(1):156--170.

\bibitem[{Goodson} et~al., 1997]{good97}
{Goodson}, A.~P., {Winglee}, R.~M., and {Bo{\"o}hm}, K.-H. (1997).
\newblock {Time-dependent Accretion by Magnetic Young Stellar Objects as a
  Launching Mechanism for Stellar Jets}.
\newblock {\em \apj}, 489:199--+.

\bibitem[{Gravity Collaboration} et~al., 2017]{2017A&A...608A..78G}
{Gravity Collaboration}, {Garcia Lopez}, R., {Perraut}, K., {Caratti O
  Garatti}, A., {Lazareff}, B., {Sanchez-Bermudez}, J., {Benisty}, M.,
  {Dougados}, C., {Labadie}, L., {Brandner}, W., {Garcia}, P.~J.~V., {Henning},
  T., {Ray}, T.~P., {Abuter}, R., {Amorim}, A., {Anugu}, N., {Berger}, J.~P.,
  {Bonnet}, H., {Buron}, A., {Caselli}, P., {Cl{\'e}net}, Y., {Coud{\'e} Du
  Foresto}, V., {de Wit}, W., {Deen}, C., {Delplancke-Str{\"o}bele}, F.,
  {Dexter}, J., {Eckart}, A., {Eisenhauer}, F., {Garcia Dabo}, C.~E.,
  {Gendron}, E., {Genzel}, R., {Gillessen}, S., {Haubois}, X., {Haug}, M.,
  {Haussmann}, F., {Hippler}, S., {Hubert}, Z., {Hummel}, C.~A., {Horrobin},
  M., {Jocou}, L., {Kellner}, S., {Kervella}, P., {Kulas}, M., {Kolb}, J.,
  {Lacour}, S., {Le Bouquin}, J.~B., {L{\'e}na}, P., {Lippa}, M., {M{\'e}rand},
  A., {M{\"u}ller}, E., {Ott}, T., {Panduro}, J., {Paumard}, T., {Perrin}, G.,
  {Pfuhl}, O., {Ramirez}, A., {Rau}, C., {Rohloff}, R.~R., {Rousset}, G.,
  {Scheithauer}, S., {Sch{\"o}ller}, M., {Straubmeier}, C., {Sturm}, E., {Thi},
  W.~F., {van Dishoeck}, E., {Vincent}, F., {Waisberg}, I., {Wank}, I.,
  {Wieprecht}, E., {Wiest}, M., {Wiezorrek}, E., {Woillez}, J., {Yazici}, S.,
  and {Zins}, G. (2017).
\newblock {The wind and the magnetospheric accretion onto the T Tauri star S
  Coronae Australis at sub-au resolution}.
\newblock {\em \aap}, 608:A78.

\bibitem[{Gressel} et~al., 2015a]{gres15}
{Gressel}, O., {Turner}, N.~J., {Nelson}, R.~P., and {McNally}, C.~P. (2015a).
\newblock {Global Simulations of Protoplanetary Disks With Ohmic Resistivity
  and Ambipolar Diffusion}.
\newblock {\em \apj}, 801:84.

\bibitem[{Gressel} et~al., 2015b]{gress15}
{Gressel}, O., {Turner}, N.~J., {Nelson}, R.~P., and {McNally}, C.~P. (2015b).
\newblock {Global Simulations of Protoplanetary Disks With Ohmic Resistivity
  and Ambipolar Diffusion}.
\newblock {\em \apj}, 801:84.

\bibitem[{G{\"u}del} et~al., 2005]{gued05}
{G{\"u}del}, M., {Skinner}, S.~L., {Briggs}, K.~R., {Audard}, M., {Arzner}, K.,
  and {Telleschi}, A. (2005).
\newblock {Evidence for an X-Ray Jet in DG Tauri A?}
\newblock {\em \apjl}, 626:L53--L56.

\bibitem[{Gueth} and {Guilloteau}, 1999]{1999A&A...343..571G}
{Gueth}, F. and {Guilloteau}, S. (1999).
\newblock {The jet-driven molecular outflow of HH 211}.
\newblock {\em \aap}, 343:571--584.

\bibitem[{Haro}, 1952]{1952ApJ...115..572H}
{Haro}, G. (1952).
\newblock {Herbig's Nebulous Objects Near NGC 1999.}
\newblock {\em \apj}, 115:572.

\bibitem[{Hartigan} et~al., 1995]{1995ApJ...452..736H}
{Hartigan}, P., {Edwards}, S., and {Ghandour}, L. (1995).
\newblock {Disk Accretion and Mass Loss from Young Stars}.
\newblock {\em \apj}, 452:736.

\bibitem[{Hartigan} et~al., 2007]{2007ApJ...661..910H}
{Hartigan}, P., {Frank}, A., {Varni{\'e}re}, P., and {Blackman}, E.~G. (2007).
\newblock {Magnetic Fields in Stellar Jets}.
\newblock {\em \apj}, 661:910--918.

\bibitem[{Hartigan} and {Morse}, 2007]{hart07a}
{Hartigan}, P. and {Morse}, J. (2007).
\newblock {Collimation, Proper Motions, and Physical Conditions in the HH 30
  Jet from Hubble Space Telescope Slitless Spectroscopy}.
\newblock {\em \apj}, 660:426--440.

\bibitem[{Hartigan} et~al., 1994]{1994ApJ...436..125H}
{Hartigan}, P., {Morse}, J.~A., and {Raymond}, J. (1994).
\newblock {Mass-Loss Rates, Ionization Fractions, Shock Velocities, and
  Magnetic Fields of Stellar Jets}.
\newblock {\em \apj}, 436:125.

\bibitem[{Hartigan} and {Wright}, 2015]{2015ApJ...811...12H}
{Hartigan}, P. and {Wright}, A. (2015).
\newblock {A New Diagnostic of Magnetic Field Strengths in Radiatively-cooled
  Shocks}.
\newblock {\em \apj}, 811:12.

\bibitem[{Hartmann} et~al., 2016]{2016ARA&A..54..135H}
{Hartmann}, L., {Herczeg}, G., and {Calvet}, N. (2016).
\newblock {Accretion onto Pre-Main-Sequence Stars}.
\newblock {\em \araa}, 54:135--180.

\bibitem[{Hartmann} and {MacGregor}, 1982]{hart82a}
{Hartmann}, L. and {MacGregor}, K.~B. (1982).
\newblock {Wave-driven winds from cool stars. I - Some effects of magnetic
  field geometry}.
\newblock {\em \apj}, 257:264--268.

\bibitem[{Hawley} et~al., 1995]{hawl95}
{Hawley}, J.~F., {Gammie}, C.~F., and {Balbus}, S.~A. (1995).
\newblock {Local Three-dimensional Magnetohydrodynamic Simulations of Accretion
  Disks}.
\newblock {\em \apj}, 440:742.

\bibitem[{Herbig}, 1951]{1951ApJ...113..697H}
{Herbig}, G.~H. (1951).
\newblock {The Spectra of Two Nebulous Objects Near NGC 1999.}
\newblock {\em \apj}, 113:697--699.

\bibitem[{Hirota} et~al., 2017]{2017NatAs...1E.146H}
{Hirota}, T., {Machida}, M.~N., {Matsushita}, Y., {Motogi}, K., {Matsumoto},
  N., {Kim}, M.~K., {Burns}, R.~A., and {Honma}, M. (2017).
\newblock {Disk-driven rotating bipolar outflow in Orion Source I}.
\newblock {\em Nature Astronomy}, 1:0146.

\bibitem[{Hussain} and {Alecian}, 2014]{2014IAUS..302...25H}
{Hussain}, G.~A.~J. and {Alecian}, E. (2014).
\newblock {The role of magnetic fields in pre-main sequence stars}.
\newblock In {Petit}, P., {Jardine}, M., and {Spruit}, H.~C., editors, {\em
  Magnetic Fields throughout Stellar Evolution}, volume 302 of {\em IAU
  Symposium}, pages 25--37.

\bibitem[{Jacquemin-Ide} et~al., 2019]{2019MNRAS.490.3112J}
{Jacquemin-Ide}, J., {Ferreira}, J., and {Lesur}, G. (2019).
\newblock {Magnetically driven jets and winds from weakly magnetized accretion
  discs}.
\newblock {\em \mnras}, 490(3):3112--3133.

\bibitem[{Jhan} and {Lee}, 2016]{2016ApJ...816...32J}
{Jhan}, K.-S. and {Lee}, C.-F. (2016).
\newblock {A Multi-epoch SMA Study of the HH 211 Protostellar Jet: Jet Motion
  and Knot Formation}.
\newblock {\em \apj}, 816(1):32.

\bibitem[{Jim{\'e}nez-Serra} et~al., 2013]{2013ApJ...764L...4J}
{Jim{\'e}nez-Serra}, I., {B{\'a}ez-Rubio}, A., {Rivilla}, V.~M.,
  {Mart{\'\i}n-Pintado}, J., {Zhang}, Q., {Dierickx}, M., and {Patel}, N.
  (2013).
\newblock {A New Radio Recombination Line Maser Object toward the MonR2 H II
  Region}.
\newblock {\em \apjl}, 764(1):L4.

\bibitem[{Koenigl}, 1991]{1991ApJ...370L..39K}
{Koenigl}, A. (1991).
\newblock {Disk Accretion onto Magnetic T Tauri Stars}.
\newblock {\em \apjl}, 370:L39.

\bibitem[{Konigl}, 1989]{koni89}
{Konigl}, A. (1989).
\newblock {Self-similar models of magnetized accretion disks}.
\newblock {\em \apj}, 342:208--223.

\bibitem[{Krasnopolsky} et~al., 1999]{kras99}
{Krasnopolsky}, R., {Li}, Z.-Y., and {Blandford}, R. (1999).
\newblock {Magnetocentrifugal Launching of Jets from Accretion Disks. I. Cold
  Axisymmetric Flows}.
\newblock {\em \apj}, 526:631--642.

\bibitem[{Krasnopolsky} et~al., 2003]{kras03}
{Krasnopolsky}, R., {Li}, Z.-Y., and {Blandford}, R.~D. (2003).
\newblock {Magnetocentrifugal Launching of Jets from Accretion Disks. II. Inner
  Disk-driven Winds}.
\newblock {\em \apj}, 595:631--642.

\bibitem[{Krumholz}, 2015]{2015arXiv151103457K}
{Krumholz}, M.~R. (2015).
\newblock {Notes on Star Formation}.
\newblock {\em arXiv e-prints}, page arXiv:1511.03457.

\bibitem[{Kwan}, 1997]{kwan97}
{Kwan}, J. (1997).
\newblock {Warm Disk Coronae in Classical T Tauri Stars}.
\newblock {\em \apj}, 489:284.

\bibitem[{Kwan} and {Tademaru}, 1995]{kwan95}
{Kwan}, J. and {Tademaru}, E. (1995).
\newblock {Disk Winds from T Tauri Stars}.
\newblock {\em \apj}, 454:382.

\bibitem[{Labdon} et~al., 2019]{2019A&A...627A..36L}
{Labdon}, A., {Kraus}, S., {Davies}, C.~L., {Kreplin}, A., {Kluska}, J.,
  {Harries}, T.~J., {Monnier}, J.~D., {ten Brummelaar}, T., {Baron}, F.,
  {Millan-Gabet}, R., {Kloppenborg}, B., {Eisner}, J., {Sturmann}, J., and
  {Sturmann}, L. (2019).
\newblock {Dusty disk winds at the sublimation rim of the highly inclined, low
  mass young stellar object SU Aurigae}.
\newblock {\em \aap}, 627:A36.

\bibitem[{Lada} and {Wilking}, 1984]{1984ApJ...287..610L}
{Lada}, C.~J. and {Wilking}, B.~A. (1984).
\newblock {The nature of the embedded population in the rho Ophiuchi dark cloud
  : mid-infrared observations.}
\newblock {\em \apj}, 287:610--621.

\bibitem[{Lee} et~al., 2017]{2017NatAs...1E.152L}
{Lee}, C.-F., {Ho}, P.~T.~P., {Li}, Z.-Y., {Hirano}, N., {Zhang}, Q., and
  {Shang}, H. (2017).
\newblock {A rotating protostellar jet launched from the innermost disk of HH
  212}.
\newblock {\em Nature Astronomy}, 1:0152.

\bibitem[{Lee} et~al., 2018]{2018NatCo...9.4636L}
{Lee}, C.-F., {Hwang}, H.-C., {Ching}, T.-C., {Hirano}, N., {Lai}, S.-P.,
  {Rao}, R., and {Ho}, P. T.~P. (2018).
\newblock {Unveiling a magnetized jet from a low-mass protostar}.
\newblock {\em Nature Communications}, 9:4636.

\bibitem[{Lesur} et~al., 2013]{lesu13}
{Lesur}, G., {Ferreira}, J., and {Ogilvie}, G.~I. (2013).
\newblock {The magnetorotational instability as a jet launching mechanism}.
\newblock {\em \aap}, 550:A61.

\bibitem[{Lesur} et~al., 2014]{lesu14}
{Lesur}, G., {Kunz}, M.~W., and {Fromang}, S. (2014).
\newblock {Thanatology in protoplanetary discs. The combined influence of
  Ohmic, Hall, and ambipolar diffusion on dead zones}.
\newblock {\em \aap}, 566:A56.

\bibitem[{Li}, 1995]{li95}
{Li}, Z.-Y. (1995).
\newblock {Magnetohydrodynamic disk-wind connection: Self-similar solutions}.
\newblock {\em \apj}, 444:848--860.

\bibitem[{Li}, 1996]{li96a}
{Li}, Z.-Y. (1996).
\newblock {Magnetohydrodynamic Disk-Wind Connection: Magnetocentrifugal Winds
  from Ambipolar Diffusion-dominated Accretion Disks}.
\newblock {\em \apj}, 465:855--+.

\bibitem[{Lima} et~al., 2010]{2010A&A...522A.104L}
{Lima}, G.~H.~R.~A., {Alencar}, S.~H.~P., {Calvet}, N., {Hartmann}, L., and
  {Muzerolle}, J. (2010).
\newblock {Modeling the H{\ensuremath{\alpha}} line emission around classical T
  Tauri stars using magnetospheric accretion and disk wind models}.
\newblock {\em \aap}, 522:A104.

\bibitem[{Louvet} et~al., 2016]{2016A&A...596A..88L}
{Louvet}, F., {Dougados}, C., {Cabrit}, S., {Hales}, A., {Pinte}, C.,
  {M{\'e}nard}, F., {Bacciotti}, F., {Coffey}, D., {Mardones}, D., {Bronfman},
  L., and {Gueth}, F. (2016).
\newblock {ALMA observations of the <ASTROBJ>Th 28</ASTROBJ> protostellar disk.
  A new example of counter-rotation between disk and optical jet}.
\newblock {\em \aap}, 596:A88.

\bibitem[{Louvet} et~al., 2018]{2018A&A...618A.120L}
{Louvet}, F., {Dougados}, C., {Cabrit}, S., {Mardones}, D., {M{\'e}nard}, F.,
  {Tabone}, B., {Pinte}, C., and {Dent}, W.~R.~F. (2018).
\newblock {The HH30 edge-on T Tauri star. A rotating and precessing monopolar
  outflow scrutinized by ALMA}.
\newblock {\em \aap}, 618:A120.

\bibitem[{Matt} and {Pudritz}, 2005]{matt05b}
{Matt}, S. and {Pudritz}, R.~E. (2005).
\newblock {Accretion-powered Stellar Winds as a Solution to the Stellar Angular
  Momentum Problem}.
\newblock {\em \apjl}, 632:L135--L138.

\bibitem[{Matt} et~al., 2012]{matt12a}
{Matt}, S.~P., {Pinz{\'o}n}, G., {Greene}, T.~P., and {Pudritz}, R.~E. (2012).
\newblock {Spin Evolution of Accreting Young Stars. II. Effect of
  Accretion-powered Stellar Winds}.
\newblock {\em \apj}, 745:101.

\bibitem[{McCaughrean} et~al., 1994]{1994ApJ...436L.189M}
{McCaughrean}, M.~J., {Rayner}, J.~T., and {Zinnecker}, H. (1994).
\newblock {Discovery of a Molecular Hydrogen Jet near IC 348}.
\newblock {\em \apjl}, 436:L189.

\bibitem[{Meliani} et~al., 2006]{meli06b}
{Meliani}, Z., {Casse}, F., and {Sauty}, C. (2006).
\newblock {Two-component magnetohydrodynamical outflows around young stellar
  objects. Interplay between stellar magnetospheric winds and disc-driven
  jets}.
\newblock {\em \aap}, 460:1--14.

\bibitem[{Melnikov} et~al., 2009]{meln09}
{Melnikov}, S.~Y., {Eisl{\"o}ffel}, J., {Bacciotti}, F., {Woitas}, J., and
  {Ray}, T.~P. (2009).
\newblock {HST/STIS observations of the RW Aurigae bipolar jet: mapping the
  physical parameters close to the source}.
\newblock {\em \aap}, 506:763--777.

\bibitem[{Mestel} and {Spruit}, 1987]{mest87}
{Mestel}, L. and {Spruit}, H.~C. (1987).
\newblock {On magnetic braking of late-type stars}.
\newblock {\em \mnras}, 226:57--66.

\bibitem[{Murphy} et~al., 2010]{murp10}
{Murphy}, G.~C., {Ferreira}, J., and {Zanni}, C. (2010).
\newblock {Large scale magnetic fields in viscous resistive accretion disks. I.
  Ejection from weakly magnetized disks}.
\newblock {\em \aap}, 512:82.

\bibitem[{Najita} and {Bergin}, 2018]{2018ApJ...864..168N}
{Najita}, J.~R. and {Bergin}, E.~A. (2018).
\newblock {Protoplanetary Disk Sizes and Angular Momentum Transport}.
\newblock {\em \apj}, 864(2):168.

\bibitem[{Natta} et~al., 2004]{2004A&A...424..603N}
{Natta}, A., {Testi}, L., {Muzerolle}, J., {Randich}, S., {Comer{\'o}n}, F.,
  and {Persi}, P. (2004).
\newblock {Accretion in brown dwarfs: An infrared view}.
\newblock {\em \aap}, 424:603--612.

\bibitem[{Nisini} et~al., 2007]{2007A&A...462..163N}
{Nisini}, B., {Codella}, C., {Giannini}, T., {Santiago Garcia}, J., {Richer},
  J.~S., {Bachiller}, R., and {Tafalla}, M. (2007).
\newblock {Warm SiO gas in molecular bullets associated with protostellar
  outflows}.
\newblock {\em \aap}, 462(1):163--172.

\bibitem[{Ogilvie} and {Livio}, 2001]{ogil01}
{Ogilvie}, G.~I. and {Livio}, M. (2001).
\newblock {Launching of Jets and the Vertical Structure of Accretion Disks}.
\newblock {\em \apj}, 553:158--173.

\bibitem[{O'Keeffe} and {Downes}, 2014]{okee14}
{O'Keeffe}, W. and {Downes}, T.~P. (2014).
\newblock {Multifluid simulations of the magnetorotational instability in
  protostellar discs}.
\newblock {\em \mnras}, 441:571--581.

\bibitem[{Ouyed} et~al., 2003]{ouye03}
{Ouyed}, R., {Clarke}, D.~A., and {Pudritz}, R.~E. (2003).
\newblock {Three-dimensional Simulations of Jets from Keplerian Disks:
  Self-regulatory Stability}.
\newblock {\em \apj}, 582:292--319.

\bibitem[{Ouyed} and {Pudritz}, 1997a]{ouye97a}
{Ouyed}, R. and {Pudritz}, R.~E. (1997a).
\newblock {Numerical Simulations of Astrophysical Jets from Keplerian Disks. I.
  Stationary Models}.
\newblock {\em \apj}, 482:712.

\bibitem[{Ouyed} and {Pudritz}, 1997b]{ouye97b}
{Ouyed}, R. and {Pudritz}, R.~E. (1997b).
\newblock {Numerical Simulations of Astrophysical Jets from Keplerian Disks.
  II. Episodic Outflows}.
\newblock {\em \apj}, 484:794.

\bibitem[{Perraut} et~al., 2016]{2016A&A...596A..17P}
{Perraut}, K., {Dougados}, C., {Lima}, G.~H.~R.~A., {Benisty}, M., {Mourard},
  D., {Ligi}, R., {Nardetto}, N., {Tallon-Bosc}, I., {ten Brummelaar}, T., and
  {Farrington}, C. (2016).
\newblock {A disk wind in AB Aurigae traced with H{\ensuremath{\alpha}}
  interferometry}.
\newblock {\em \aap}, 596:A17.

\bibitem[{Podio} et~al., 2011]{podi11}
{Podio}, L., {Eisl{\"o}ffel}, J., {Melnikov}, S., {Hodapp}, K.~W., and
  {Bacciotti}, F. (2011).
\newblock {Tracing kinematical and physical asymmetries in the jet from DG
  Tauri B}.
\newblock {\em \aap}, 527:A13.

\bibitem[{Podio} et~al., 2009]{2009A&A...506..779P}
{Podio}, L., {Medves}, S., {Bacciotti}, F., {Eisl{\"o}ffel}, J., and {Ray}, T.
  (2009).
\newblock {Physical structure and dust reprocessing in a sample of HH jets}.
\newblock {\em \aap}, 506(2):779--788.

\bibitem[{Pudritz} and {Norman}, 1986]{pudr86}
{Pudritz}, R.~E. and {Norman}, C.~A. (1986).
\newblock {Bipolar hydromagnetic winds from disks around protostellar objects}.
\newblock {\em \apj}, 301:571--586.

\bibitem[{Pudritz} and {Ray}, 2019]{2019FrASS...6...54P}
{Pudritz}, R.~E. and {Ray}, T.~P. (2019).
\newblock {The Role of Magnetic Fields in Protostellar Outflows and Star
  Formation}.
\newblock {\em Frontiers in Astronomy and Space Sciences}, 6:54.

\bibitem[{Pudritz} et~al., 2006]{pudr06}
{Pudritz}, R.~E., {Rogers}, C.~S., and {Ouyed}, R. (2006).
\newblock {Controlling the collimation and rotation of hydromagnetic disc
  winds}.
\newblock {\em \mnras}, 365:1131--1148.

\bibitem[{Purser} et~al., 2016]{2016MNRAS.460.1039P}
{Purser}, S.~J.~D., {Lumsden}, S.~L., {Hoare}, M.~G., {Urquhart}, J.~S.,
  {Cunningham}, N., {Purcell}, C.~R., {Brooks}, K.~J., {Garay}, G.,
  {G{\'u}zman}, A.~E., and {Voronkov}, M.~A. (2016).
\newblock {A search for ionized jets towards massive young stellar objects}.
\newblock {\em \mnras}, 460(1):1039--1053.

\bibitem[{Qiu} et~al., 2019]{2019ApJ...871..141Q}
{Qiu}, K., {Wyrowski}, F., {Menten}, K., {Zhang}, Q., and {G{\"u}sten}, R.
  (2019).
\newblock {CO Multi-line Observations of HH 80-81: A Two-component Molecular
  Outflow Associated with the Largest Protostellar Jet in Our Galaxy}.
\newblock {\em \apj}, 871(2):141.

\bibitem[{Raga} and {Kofman}, 1992]{1992ApJ...386..222R}
{Raga}, A.~C. and {Kofman}, L. (1992).
\newblock {Knots in Stellar Jets from Time-dependent Sources}.
\newblock {\em \apj}, 386:222.

\bibitem[{Ramsey} and {Clarke}, 2011]{rams11}
{Ramsey}, J.~P. and {Clarke}, D.~A. (2011).
\newblock {Simulating Protostellar Jets Simultaneously at Launching and
  Observational Scales}.
\newblock {\em \apjl}, 728:L11.

\bibitem[{Rees}, 1978]{1978MNRAS.184P..61R}
{Rees}, M.~J. (1978).
\newblock {The M87 jet: internal shocks in a plasma beam?}
\newblock {\em \mnras}, 184:61P--65P.

\bibitem[{Reynolds}, 1986]{1986ApJ...304..713R}
{Reynolds}, S.~P. (1986).
\newblock {Continuum Spectra of Collimated, Ionized Stellar Winds}.
\newblock {\em \apj}, 304:713.

\bibitem[{Richards} et~al., 2018]{2018JAI.....740015R}
{Richards}, S.~N., {Moseley}, S.~H., {Stacey}, G., {Greenhouse}, M., {Kutyrev},
  A., {Arendt}, R., {Atanasoff}, H., {Banks}, S., {Brekosky}, R.~P., {Brown},
  A.-D., {Bulcha}, B., {Cazeau}, T., {Choi}, M., {Colazo}, F., {Engler}, C.,
  {Hadjimichael}, T., {Hays-Wehle}, J., {Henderson}, C., {Hsieh}, W.-T.,
  {Huang}, J., {Jenstrom}, I., {Kellogg}, J., {Kimball}, M., {Kov{\'a}cs}, A.,
  {Leiter}, S., {Maher}, S., {McMurray}, R., {Melnick}, G.~J., {Mentzell}, E.,
  {Mikula}, V., {Miller}, T.~M., {Nagler}, P., {Nikola}, T., {Oxborrow}, J.,
  {Pontoppidan}, K.~M., {Rangwala}, N., {Rhodes}, A., {Roberge}, A., {Rosner},
  S., {Rostem}, K., {Rustemeyer}, N., {Sharp}, E., {Sparr}, L., {Stevanovic},
  D., {Taraschi}, P., {Temi}, P., {Vacca}, W.~D., {de Lorenzo}, J. V.~H.,
  {Wohler}, B., {Wollack}, E.~J., and {Wilks}, S. (2018).
\newblock {SOFIA-HIRMES: Looking Forward to the HIgh-Resolution Mid-infrarEd
  Spectrometer}.
\newblock {\em Journal of Astronomical Instrumentation}, 7(4):1840015.

\bibitem[{Richer} et~al., 2000]{2000prpl.conf..867R}
{Richer}, J.~S., {Shepherd}, D.~S., {Cabrit}, S., {Bachiller}, R., and
  {Churchwell}, E. (2000).
\newblock {Molecular Outflows from Young Stellar Objects}.
\newblock In {Mannings}, V., {Boss}, A.~P., and {Russell}, S.~S., editors, {\em
  Protostars and Planets IV}, page 867.

\bibitem[{Rodgers-Lee} et~al., 2017]{2017MNRAS.472...26R}
{Rodgers-Lee}, D., {Taylor}, A.~M., {Ray}, T.~P., and {Downes}, T.~P. (2017).
\newblock {The ionizing effect of low-energy cosmic rays from a class II object
  on its protoplanetary disc}.
\newblock {\em \mnras}, 472(1):26--38.

\bibitem[{Rodriguez} et~al., 1989]{1989ApJ...346L..85R}
{Rodriguez}, L.~F., {Curiel}, S., {Moran}, J.~M., {Mirabel}, I.~F., {Roth}, M.,
  and {Garay}, G. (1989).
\newblock {Large Proper Motions in the Remarkable Triple Radio Source in
  Serpens}.
\newblock {\em \apjl}, 346:L85.

\bibitem[{Rodr{\'\i}guez-Kamenetzky} et~al., 2019]{2019MNRAS.482.4687R}
{Rodr{\'\i}guez-Kamenetzky}, A., {Carrasco-Gonz{\'a}lez}, C.,
  {Gonz{\'a}lez-Mart{\'\i}n}, O., {Araudo}, A.~T., {Rodr{\'\i}guez}, L.~F.,
  {Vig}, S., and {Hofner}, P. (2019).
\newblock {Particle acceleration in the Herbig-Haro objects HH 80 and HH 81}.
\newblock {\em \mnras}, 482(4):4687--4696.

\bibitem[{Romanova} et~al., 2002]{roma02}
{Romanova}, M.~M., {Ustyugova}, G.~V., {Koldoba}, A.~V., and {Lovelace},
  R.~V.~E. (2002).
\newblock {Magnetohydrodynamic Simulations of Disk-Magnetized Star Interactions
  in the Quiescent Regime: Funnel Flows and Angular Momentum Transport}.
\newblock {\em \apj}, 578:420--438.

\bibitem[{Romanova} et~al., 2009]{roma09}
{Romanova}, M.~M., {Ustyugova}, G.~V., {Koldoba}, A.~V., and {Lovelace},
  R.~V.~E. (2009).
\newblock {Launching of conical winds and axial jets from the
  disc-magnetosphere boundary: axisymmetric and 3D simulations}.
\newblock {\em \mnras}, 399:1802--1828.

\bibitem[{Romanova} et~al., 2012]{roma12}
{Romanova}, M.~M., {Ustyugova}, G.~V., {Koldoba}, A.~V., and {Lovelace},
  R.~V.~E. (2012).
\newblock {MRI-driven accretion on to magnetized stars: global 3D MHD
  simulations of magnetospheric and boundary layer regimes}.
\newblock {\em \mnras}, 421:63--77.

\bibitem[{Salmeron} et~al., 2011]{salm11}
{Salmeron}, R., {K{\"o}nigl}, A., and {Wardle}, M. (2011).
\newblock {Wind-driving protostellar accretion discs - II. Numerical method and
  illustrative solutions}.
\newblock {\em \mnras}, 412:1162--1180.

\bibitem[{Scepi} et~al., 2018]{scep18}
{Scepi}, N., {Lesur}, G., {Dubus}, G., and {Flock}, M. (2018).
\newblock {Turbulent and wind-driven accretion in dwarf novae threaded by a
  large-scale magnetic field}.
\newblock {\em \aap}, 620:A49.

\bibitem[{Schatzman}, 1962]{scha62}
{Schatzman}, E. (1962).
\newblock {A theory of the role of magnetic activity during star formation}.
\newblock {\em Annales d'Astrophysique}, 25:18.

\bibitem[{Schwartz}, 1977]{1977ApJ...212L..25S}
{Schwartz}, R.~D. (1977).
\newblock {Evidence of star formation triggered by expansion of the Gum
  Nebula.}
\newblock {\em \apjl}, 212:L25--L26.

\bibitem[{Shakura} and {Sunyaev}, 1973]{shak73}
{Shakura}, N.~I. and {Sunyaev}, R.~A. (1973).
\newblock {Black holes in binary systems. Observational appearance.}
\newblock {\em \aap}, 24:337--355.

\bibitem[{Sheikhnezami} and {Fendt}, 2015]{2015ApJ...814..113S}
{Sheikhnezami}, S. and {Fendt}, C. (2015).
\newblock {Wobbling and Precessing Jets from Warped Disks in Binary Systems}.
\newblock {\em \apj}, 814(2):113.

\bibitem[{Sheikhnezami} and {Fendt}, 2018]{2018ApJ...861...11S}
{Sheikhnezami}, S. and {Fendt}, C. (2018).
\newblock {Long-term Simulation of MHD Jet Launching in an Orbiting Star-Disk
  System}.
\newblock {\em \apj}, 861(1):11.

\bibitem[{Sheikhnezami} et~al., 2012]{shei12}
{Sheikhnezami}, S., {Fendt}, C., {Porth}, O., {Vaidya}, B., and {Ghanbari}, J.
  (2012).
\newblock {Bipolar Jets Launched from Magnetically Diffusive Accretion Disks.
  I. Ejection Efficiency versus Field Strength and Diffusivity}.
\newblock {\em \apj}, 757:65.

\bibitem[{Shu} et~al., 1994]{shu94a}
{Shu}, F., {Najita}, J., {Ostriker}, E., {Wilkin}, F., {Ruden}, S., and
  {Lizano}, S. (1994).
\newblock {Magnetocentrifugally driven flows from young stars and disks. 1: A
  generalized model}.
\newblock {\em \apj}, 429:781--796.

\bibitem[{Shu} et~al., 1988]{shu88}
{Shu}, F.~H., {Lizano}, S., {Ruden}, S.~P., and {Najita}, J. (1988).
\newblock {Mass loss from rapidly rotating magnetic protostars}.
\newblock {\em \apjl}, 328:L19--L23.

\bibitem[{Skinner} et~al., 2011]{skin11}
{Skinner}, S.~L., {Audard}, M., and {G{\"u}del}, M. (2011).
\newblock {Chandra Evidence for Extended X-Ray Structure in RY Tau}.
\newblock {\em \apj}, 737:19.

\bibitem[{Snell} et~al., 1985]{1985ApJ...290..587S}
{Snell}, R.~L., {Bally}, J., {Strom}, S.~E., and {Strom}, K.~M. (1985).
\newblock {Radio and optical observations of the jets L 1551 IRS 5.}
\newblock {\em \apj}, 290:587--595.

\bibitem[{Snell} et~al., 1980]{1980ApJ...239L..17S}
{Snell}, R.~L., {Loren}, R.~B., and {Plambeck}, R.~L. (1980).
\newblock {Observations of CO in L 1551 : evidence for stellar wind driven
  shocks.}
\newblock {\em \apjl}, 239:L17--L22.

\bibitem[{Staff} et~al., 2015]{staf15}
{Staff}, J.~E., {Koning}, N., {Ouyed}, R., {Thompson}, A., and {Pudritz}, R.~E.
  (2015).
\newblock {Hubble Space Telescope scale 3D simulations of MHD disc winds: a
  rotating two-component jet structure}.
\newblock {\em \mnras}, 446(4):3975--3991.

\bibitem[{Staff} et~al., 2010]{staf10}
{Staff}, J.~E., {Niebergal}, B.~P., {Ouyed}, R., {Pudritz}, R.~E., and {Cai},
  K. (2010).
\newblock {Confronting Three-dimensional Time-dependent Jet Simulations with
  Hubble Space Telescope Observations}.
\newblock {\em \apj}, 722:1325--1332.

\bibitem[{Stepanovs} and {Fendt}, 2016]{step16}
{Stepanovs}, D. and {Fendt}, C. (2016).
\newblock {An Extensive Numerical Survey of the Correlation Between Outflow
  Dynamics and Accretion Disk Magnetization}.
\newblock {\em \apj}, 825:14.

\bibitem[{Stojimirovi{\'c}} et~al., 2006]{2006ApJ...649..280S}
{Stojimirovi{\'c}}, I., {Narayanan}, G., {Snell}, R.~L., and {Bally}, J.
  (2006).
\newblock {Entrainment Mechanisms for Outflows in the L1551 Star-forming
  Region}.
\newblock {\em \apj}, 649(1):280--298.

\bibitem[{Stone} et~al., 1996]{1996ApJ...463..656S}
{Stone}, J.~M., {Hawley}, J.~F., {Gammie}, C.~F., and {Balbus}, S.~A. (1996).
\newblock {Three-dimensional Magnetohydrodynamical Simulations of Vertically
  Stratified Accretion Disks}.
\newblock {\em \apj}, 463:656.

\bibitem[{Suriano} et~al., 2018]{suri18}
{Suriano}, S.~S., {Li}, Z.-Y., {Krasnopolsky}, R., and {Shang}, H. (2018).
\newblock {The formation of rings and gaps in magnetically coupled disc-wind
  systems: ambipolar diffusion and reconnection}.
\newblock {\em \mnras}, 477:1239--1257.

\bibitem[{Suzuki} and {Inutsuka}, 2009]{suzu09}
{Suzuki}, T.~K. and {Inutsuka}, S.-i. (2009).
\newblock {Disk Winds Driven by Magnetorotational Instability and Dispersal of
  Protoplanetary Disks}.
\newblock {\em \apjl}, 691:L49--L54.

\bibitem[{Suzuki} and {Inutsuka}, 2014]{suzu14}
{Suzuki}, T.~K. and {Inutsuka}, S.-i. (2014).
\newblock {Magnetohydrodynamic Simulations of Global Accretion Disks with
  Vertical Magnetic Fields}.
\newblock {\em \apj}, 784:121.

\bibitem[{Takeuchi} and {Okuzumi}, 2014]{take14}
{Takeuchi}, T. and {Okuzumi}, S. (2014).
\newblock {Radial Transport of Large-scale Magnetic Fields in Accretion Disks.
  II. Relaxation to Steady States}.
\newblock {\em \apj}, 797:132.

\bibitem[{Tappe} et~al., 2012]{2012ApJ...751....9T}
{Tappe}, A., {Forbrich}, J., {Mart{\'\i}n}, S., {Yuan}, Y., and {Lada}, C.~J.
  (2012).
\newblock {The Anatomy of the Young Protostellar Outflow HH 211}.
\newblock {\em \apj}, 751(1):9.

\bibitem[{Tychoniec} et~al., 2018]{2018ApJS..238...19T}
{Tychoniec}, {\L}., {Tobin}, J.~J., {Karska}, A., {Chand ler}, C., {Dunham},
  M.~M., {Harris}, R.~J., {Kratter}, K.~M., {Li}, Z.-Y., {Looney}, L.~W.,
  {Melis}, C., {P{\'e}rez}, L.~M., {Sadavoy}, S.~I., {Segura-Cox}, D., and {van
  Dishoeck}, E.~F. (2018).
\newblock {The VLA Nascent Disk and Multiplicity Survey of Perseus Protostars
  (VANDAM). IV. Free-Free Emission from Protostars: Links to Infrared
  Properties, Outflow Tracers, and Protostellar Disk Masses}.
\newblock {\em \apjs}, 238(2):19.

\bibitem[{Tzeferacos} et~al., 2009]{tzef09}
{Tzeferacos}, P., {Ferrari}, A., {Mignone}, A., {Zanni}, C., {Bodo}, G., and
  {Massaglia}, S. (2009).
\newblock {On the magnetization of jet-launching discs}.
\newblock {\em \mnras}, 400:820--834.

\bibitem[{Tzeferacos} et~al., 2013]{tzef13}
{Tzeferacos}, P., {Ferrari}, A., {Mignone}, A., {Zanni}, C., {Bodo}, G., and
  {Massaglia}, S. (2013).
\newblock {Effects of entropy generation in jet-launching discs}.
\newblock {\em \mnras}, 428:3151--3163.

\bibitem[{Venuti} et~al., 2015]{2015A&A...581A..66V}
{Venuti}, L., {Bouvier}, J., {Irwin}, J., {Stauffer}, J.~R., {Hillenbrand},
  L.~A., {Rebull}, L.~M., {Cody}, A.~M., {Alencar}, S.~H.~P., {Micela}, G.,
  {Flaccomio}, E., and {Peres}, G. (2015).
\newblock {UV variability and accretion dynamics in the young open cluster NGC
  2264}.
\newblock {\em \aap}, 581:A66.

\bibitem[{Wang} et~al., 2019]{wang18}
{Wang}, L., {Bai}, X.-N., and {Goodman}, J. (2019).
\newblock {Global Simulations of Protoplanetary Disk Outflows with Coupled
  Non-ideal Magnetohydrodynamics and Consistent Thermochemistry}.
\newblock {\em \apj}, 874(1):90.

\bibitem[{Wardle} and {K{\"o}nigl}, 1993]{ward93}
{Wardle}, M. and {K{\"o}nigl}, A. (1993).
\newblock {The structure of protostellar accretion disks and the origin of
  bipolar flows}.
\newblock {\em \apj}, 410:218--238.

\bibitem[{Watson} et~al., 2016]{2016ApJ...828...52W}
{Watson}, D.~M., {Calvet}, N.~P., {Fischer}, W.~J., {Forrest}, W.~J., {Manoj},
  P., {Megeath}, S.~T., {Melnick}, G.~J., {Najita}, J., {Neufeld}, D.~A.,
  {Sheehan}, P.~D., {Stutz}, A.~M., and {Tobin}, J.~J. (2016).
\newblock {Evolution of Mass Outflow in Protostars}.
\newblock {\em \apj}, 828:52.

\bibitem[{Weigelt} et~al., 2011]{2011A&A...527A.103W}
{Weigelt}, G., {Grinin}, V.~P., {Groh}, J.~H., {Hofmann}, K.~H., {Kraus}, S.,
  {Miroshnichenko}, A.~S., {Schertl}, D., {Tambovtseva}, L.~V., {Benisty}, M.,
  {Driebe}, T., {Lagarde}, S., {Malbet}, F., {Meilland}, A., {Petrov}, R., and
  {Tatulli}, E. (2011).
\newblock {VLTI/AMBER spectro-interferometry of the Herbig Be star MWC 297 with
  spectral resolution 12 000}.
\newblock {\em \aap}, 527:A103.

\bibitem[{Whelan} et~al., 2004]{2004A&A...417..247W}
{Whelan}, E.~T., {Ray}, T.~P., and {Davis}, C.~J. (2004).
\newblock {Paschen beta emission as a tracer of outflow activity from T-Tauri
  stars, as compared to optical forbidden emission}.
\newblock {\em \aap}, 417:247--261.

\bibitem[{Yan} et~al., 2019]{2019arXiv190810994Y}
{Yan}, D., {Zhou}, J., and {Zhang}, P. (2019).
\newblock {Detection of $\gamma$-rays from the Protostellar Jet in the HH 80-81
  System}.
\newblock {\em arXiv e-prints}, page arXiv:1908.10994.

\bibitem[{Zanni} et~al., 2007]{zann07}
{Zanni}, C., {Ferrari}, A., {Rosner}, R., {Bodo}, G., and {Massaglia}, S.
  (2007).
\newblock {MHD simulations of jet acceleration from Keplerian accretion disks:
  the effects of disk resistivity}.
\newblock {\em \aap}, 469:811.

\bibitem[{Zanni} and {Ferreira}, 2009]{zann09}
{Zanni}, C. and {Ferreira}, J. (2009).
\newblock {MHD simulations of accretion onto a dipolar magnetosphere. I.
  Accretion curtains and the disk-locking paradigm}.
\newblock {\em \aap}, 508:1117--1133.

\bibitem[{Zanni} and {Ferreira}, 2011]{zann11}
{Zanni}, C. and {Ferreira}, J. (2011).
\newblock {Observational Limits on the Spin-down Torque of Accretion Powered
  Stellar Winds}.
\newblock {\em \apjl}, 727:L22.

\bibitem[{Zanni} and {Ferreira}, 2013]{zann13}
{Zanni}, C. and {Ferreira}, J. (2013).
\newblock {MHD simulations of accretion onto a dipolar magnetosphere. II.
  Magnetospheric ejections and stellar spin-down}.
\newblock {\em \aap}, 550:A99.

\bibitem[{Zhang} et~al., 2017]{2017ApJ...837...53Z}
{Zhang}, Q., {Claus}, B., {Watson}, L., and {Moran}, J. (2017).
\newblock {Angular Momentum in Disk Wind Revealed in the Young Star MWC 349A}.
\newblock {\em \apj}, 837(1):53.

\bibitem[{Zhu} and {Stone}, 2018]{zhu18}
{Zhu}, Z. and {Stone}, J.~M. (2018).
\newblock {Global Evolution of an Accretion Disk with a Net Vertical Field:
  Coronal Accretion, Flux Transport, and Disk Winds}.
\newblock {\em \apj}, 857:34.

\end{thebibliography}

\end{document}